\documentclass[fleqn,usenatbib]{mnras}

\usepackage{newtxtext,newtxmath}

\usepackage[T1]{fontenc}
\usepackage{ae,aecompl}
\usepackage{multicol, blindtext}
\usepackage{multirow}
\usepackage{pdflscape}
\usepackage{csquotes}
\usepackage{siunitx}

\DeclareRobustCommand{\VAN}[3]{#2}
\let\VANthebibliography\thebibliography
\def\thebibliography{\DeclareRobustCommand{\VAN}[3]{##3}\VANthebibliography}

\usepackage{graphicx}	
\usepackage{amsmath}	
\DeclareUnicodeCharacter{2212}{-}

\title[Luminous AGB Stars in M31's Satellites and Stellar Halo]{A near-IR survey of luminous asymptotic giant branch stars in the satellites and stellar halo of M31}

\author[J.~M.~Howell et al.]{
Jess~M.~Howell,$^{1}$\thanks{E-mail:jess.howell@ed.ac.uk}
Annette~M.~N.~Ferguson$^{1}$, 
Olivia~C.~Jones$^{2}$,
Mike.~J.~Irwin$^{3}$,
Maria-Rosa~L.~Cioni$^{4}$,
\newauthor Geraint~F.~Lewis$^{5}$,
Nicolas.~F.~Martin$^{6,7}$,
Alan.~W.~McConnachie$^{8}$
\\
$^{1}$Institute for Astronomy, University of Edinburgh, Royal Observatory, Blackford Hill, Edinburgh EH9 3HJ UK \\
$^{2}$UK Astronomy Technology Centre, Royal Observatory,
Blackford Hill, Edinburgh, EH9 3HJ, UK \\
$^{3}$Institute of Astronomy, University of Cambridge, Madingley Road, Cambridge CB3 0HA, UK \\
$^4$Leibniz-Institut f\"{u}r Astrophysik Potsdam, An der Sternwarte 16, D-14482 Potsdam, Germany\\
$^5$Sydney Institute for Astronomy, School of Physics A28, The University of Sydney, NSW 2006, Australia\\
$^6$Universit\'e de Strasbourg, CNRS, Observatoire astronomique de Strasbourg, UMR 7550, F-67000 Strasbourg, France\\
$^7$Max-Planck-Institut f\"{u}r Astronomie, K\"{o}nigstuhl 17, D-69117 Heidelberg, Germany \\
$^8$National Research Council Herzberg Astronomy and Astrophysics, 5071 West Saanich Road, Victoria, B.C., V8Z6M7, Canada\\
}

\date{Accepted XXX. Received YYY; in original form ZZZ}

\pubyear{2026}

\begin{document}

\label{firstpage}
\pagerange{\pageref{firstpage}--\pageref{lastpage}}
\maketitle

\begin{abstract}
We investigate the dwarf spheroidal (dSph) satellites and stellar halo of M31 using near-infrared imaging from the Wide-Field Camera on the UKIRT 3.8~m telescope. Our homogeneous analysis covers the stellar extents of 12 dSphs, as well as their surrounding halo field populations. We identify carbon-rich (C) and oxygen-rich (M) thermally-pulsing AGB stars (TP-AGBs) using colour-magnitude and colour-colour diagrams, which also allows for effective contaminant rejection. TP-AGB stars are detected in eight of the 12 dSphs, indicating small intermediate-age populations in a substantial fraction of the sample; they are also found in considerable numbers throughout most of M31's stellar halo. By combining 316 new detections with a revised classification of objects previously reported in the literature, we increase the TP-AGB candidate population in the dSphs from 82 to 359. We use the resulting dSph and halo AGB samples to compute C/M ratios and infer metallicities in the dSphs and adjacent halo fields with non-zero AGB counts. A clear metallicity gradient is observed in the M31 stellar halo out to ${\sim}150$\,kpc, consistent with that found using other tracers. Using literature star formation histories from main-sequence turn-off photometry, we find a tight correlation between the number of C stars in the dSphs and the stellar mass formed in the last 0.5--3 Gyr. This study underscores the promise of AGB stars as quantitative tracers of intermediate-age star formation, particularly in intrinsically faint systems, and provides a catalogue of candidate TP-AGB stars for follow-up.
\end{abstract}




\section{Introduction}
\label{intro}
\begin{figure}
	\centering
	\includegraphics[width=\columnwidth,keepaspectratio]{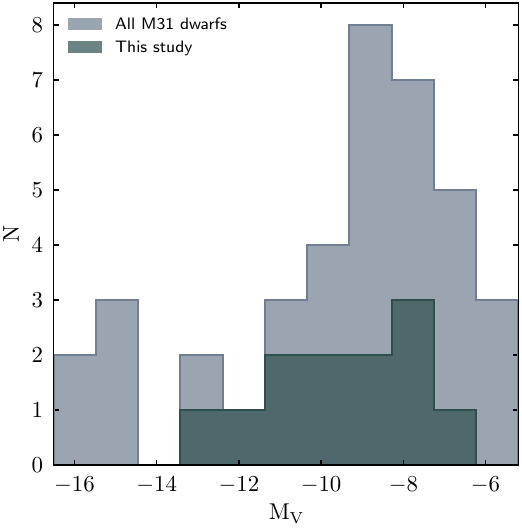}
	\caption{Light grey: luminosity distribution of all known dwarf satellites with M31 as their host, taken from \citet{Pace2025}. Dark grey: luminosity distribution of our dSph sample.}
\label{hist}
\end{figure}

Dwarf spheroidal (dSph) galaxies are typically characterised by old stellar populations and little to no interstellar material \citep[e.g.][]{Tolstoy2009}. They provide unique laboratories for investigating the processes that regulate star formation and quenching in low-mass and low-metallicity regimes, where the impacts of internal and environmental processes are most pronounced. Their evolutionary histories preserve signatures of these processes, providing insights into the evolution of satellite galaxies and the assembly of larger systems. Most Milky Way (MW) satellites quenched early and formed the bulk of their stars $\gtrsim 11$ Gyr ago \citep{Tolstoy2009, brown2014, Durbin2025}. A small number of others, such as Leo I, Carina, and Fornax, experienced prolonged star formation,  in some cases continuing to form stars until just a few hundred Myr ago \citep{RuizLara2021, Rusakov2021, deBoer2014}. On the other hand, extended star formation appears to have been more widespread in the M31 satellite population. While some M31 dSphs quenched early, like the majority of MW satellites, most continued low-level star formation until intermediate epochs. They typically show prominent early bursts of star formation followed by low-level activity until ${\sim}4$--9 Gyr ago \citep{Savino2025, Weisz2014, Skillman2017, Martin2017}, albeit with little evidence for the very recent star formation seen in some MW satellites. Extended star formation is reflected in the abundance of intermediate-age stellar populations, such as Asymptotic Giant Branch (AGB) stars. For instance, Leo I, Carina, and Fornax are all known to host carbon-rich AGB stars \citep{Groenewegen2009, Abia2008, Held2010}. Given that many M31 dSphs exhibit evidence for low-level extended star formation, it is therefore worthwhile to undertake a systematic census of AGB stars in these galaxies.

AGB stars represent a late stage of low- and intermediate-mass stellar evolution (see \citealt{Habing2004} and \citealt{Herwig2005} for reviews). Following core helium exhaustion, the star enters the early-AGB phase, where helium burning ignites in a shell beneath an already active hydrogen-burning shell. A subset of these stars will evolve into thermally-pulsing AGB stars (TP-AGBs), characterised by periodic thermonuclear runaway events (thermal pulses) driven by the unstable He-burning shell. These pulses drive deep convective mixing between the He-rich intershell region (where primary $^{12}\mathrm{C}$ is synthesised) and the surface through a process known as the third dredge-up (TDU). This can shift the surface chemistry from oxygen-rich (M stars,  $\mathrm{C}/\mathrm{O}<1$) to carbon-rich (C stars, $\mathrm{C}/\mathrm{O}>1$), depending on the star’s initial mass and metallicity. At lower metallicities, less carbon needs to be dredged to the surface for a star to reach $\mathrm{C}/\mathrm{O}>1$, and the efficiency of the TDU itself tends to increase with decreasing metallicity \citep[see e.g.][]{Karakas2002}. As a result, the ratio of observed carbon- to oxygen-rich AGB stars (the C/M ratio) is widely used as a proxy for the metallicity of a stellar population. Higher C/M ratios indicate lower metallicities, and vice versa \citep{Renzini1981, Iben1983, Cioni2003, Marigo2017}.

C/M ratios can also provide constraints on the age of a population, given the relatively narrow age range traced by C stars. For metallicities in the range of [M/H] $=−1$ to $−2$ dex, C star production peaks between ${\sim}0.5$--3 Gyr \citep{Cioni2003, Marigo2017} after the star formation event, and declines rapidly after 3 Gyr. On the other hand, M star production occurs over a longer time period, peaking between ${\sim}0.5-6$ Gyr and persisting to ${\sim}10$ Gyr \citep{Marigo2017}.

TP-AGB stars are also important in their role as prominent producers of dust and heavy elements during the final stages of their evolution, enriching the interstellar medium used in subsequent star formation. They have been shown to produce dust in as little as 30 Myr in the low-metallicity conditions of nearby dwarf galaxies \citep{McDonald2010, Boyer2017}. Furthermore, it has been shown that AGB stars are required to reproduce the Solar System abundances of carbon, nitrogen and many heavy elements produced via slow neutron capture \citep{Karakas2014, Kobayashi2020}. Nearby dwarf galaxies act as analogues to early-universe galaxies, where observations indicate that large amounts of dust are present at redshifts out to $z \sim 6$ \citep{Bertoldi2003, Algera2023}. However, the relative contributions of AGB stars to the dust and metal budgets of these galaxies remain uncertain, and it is still unclear how metallicity affects chemical yields and dust production \citep{Boyer2017, vanLoon2008, McDonald2011, Karakas2022}. Larger, homogeneous TP-AGB samples are needed to study the dust-producing phase at low metallicity, further motivating a systematic census of AGB stars in dSphs.

Stellar evolution models (e.g., \citealt{Marigo2008, Marigo2013}) predict that $>90\%$ of TP-AGB stars are brighter than the TRGB \citep{Boyer2015}. Since the TRGB can be identified observationally as the luminosity at which the RGB terminates, it therefore provides an effective threshold for identifying carbon- and oxygen-rich TP-AGB populations in resolved-star imaging of external galaxies.  By focusing on stars above this luminosity limit, we can efficiently isolate carbon- and oxygen-rich TP-AGB stars and use their relative populations as tracers of dust production, age, and metallicity across different environments.

The M31 halo has been studied in much detail in recent years thanks to deep, wide-field surveys such as the Pan-Andromeda Archaeological Survey (PAndAS; \citealt{McConnachie2018}).  This has revealed a very extended stellar halo, with its smooth metal-poor component traced out to at least ${\sim}$150 kpc \citep{Ibata2014}. These surveys have also dramatically expanded the known satellite population of M31 \citep{Martin2006,McConnachie2008,Richardson2011}. Even so, the full extent of M31’s satellite system and stellar halo has yet to be fully characterised, with new ultra-faint satellites continuing to be found \citep{Collins2022,McQuinn2023,Smith2025}. 

The early studies of M31’s satellites typically relied on small samples and focused observations \citep{Mould1990, Armandroff1993, daCosta1996, Armandroff1999, Grebel1999, daCosta2000, Harbeck2001, daCosta2002, Weisz2015, Monelli2016, Skillman2017}. More recently, a few studies have advanced this work by conducting unified analyses encompassing the majority of M31’s dSph satellite population \citep[e.g.][]{Martin2017, Savino2025}. Despite this progress, the AGB populations in the M31 dSphs are relatively understudied, with only a handful of earlier investigations examining them directly and highlighting their diagnostic potential. For example, \citet{Aaronson1985b} and \citet{Cote1999} identified C stars in the M31 dSph And\,II, demonstrating the presence of intermediate-age populations. Subsequent work expanded these efforts using targeted surveys of individual galaxies (e.g., \citealt{Harbeck2004, Kerschbaum2004, Battinelli2005, Kang2005}), often relying on optical narrow-band techniques (such as CN–TiO filters) to identify C and M stars. While these studies established that M31 satellites exhibit a wide range of C-star content, they were limited by small sample sizes, heterogeneous selection criteria, and incomplete spatial coverage. It has thus not been possible to consistently compare AGB populations across M31 dwarf satellite galaxies.

In recent years, renewed interest in this approach has emerged, with several studies demonstrating the utility of AGB stars as tracers of intermediate-age stellar populations (e.g., \citealt{Hamren2016, Navabi2021, Ren2022, Goldman2022, Jones2023, Gavetti2025}). Their importance also extends beyond the Local Group, as TP-AGB stars can contribute substantially to the integrated near-infrared light of galaxies at high redshifts \citep[e.g.,][]{Maraston2006, Melbourne2012}, a prediction recently supported by \emph{JWST}/NIRSpec observations of quiescent galaxies at $z=1$--2 \citep{Lu2025}.

\citet{Hamren2016} identified C stars in some of the satellites and halo of M31 using moderate-resolution optical spectroscopy, though with limited fields of view. \citet{Navabi2021} used long-period variable stars, the most evolved AGB and red supergiant stars, to infer the star formation history of And\,VII, while \citet{Ren2022}, \citet{Goldman2022}, and \citet{Jones2023} demonstrated the effectiveness of near-infrared (NIR) colour–magnitude selection for isolating TP-AGB populations in the M31 system. However, again, these studies typically targeted individual galaxies or a limited region of the halo.

As a result, despite the long-standing evidence of TP-AGB stars in the M31 system, a homogeneous census of these stars across the dSph population has so far been lacking. The goal of this study is to remedy this situation by conducting a uniform analysis of the dSph AGB populations using deep, wide-field NIR imaging that captures the full stellar extent of the dSphs, enabling reliable comparisons of the intermediate-age populations between satellite galaxies. Thanks to the large areal coverage of our pointings, we are also able to explore the AGB populations in the extended M31 halo using the regions adjacent to the dSphs. Section~\ref{sample} presents a brief summary of the dSphs considered, including the most up-to-date physical data and gives an overview of the data reduction process. Section~\ref{agbsel} describes the techniques used to select AGB stars and analyse the United Kingdom Infrared Telescope (UKIRT) data, and Section~\ref{discussion} presents a discussion on these populations, and the implications for the M31 dSphs and halo.

\begin{figure}

	\centering
	\includegraphics[width=\columnwidth,keepaspectratio]{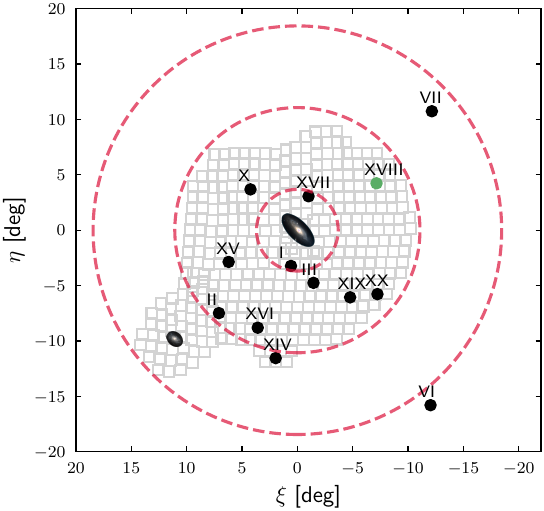}
	\caption{The distribution of the dSphs considered in this study. The full region displayed extends beyond M31's currently surveyed halo limit, which is plotted in the grey tiles (the PAndAS survey footprint \citealt{McConnachie2018}). The dashed pink circles correspond to projected radii of 50 kpc, 150 kpc, and 250 kpc from the centre of M31. And\,XVIII, highlighted in green, is used as a contamination-correction field. M31 and M33 are both shown optical from DSS2.}
\label{pandas_spatial}

\end{figure}

\begin{table*}
    \centering
    \begin{tabular}{l r r r r r r r r}
    \hline
    \hline
    dSph & $d$ (kpc) & $\langle[\mathrm{Fe}/\mathrm{H}]\rangle$ & $\epsilon = 1 - b/a$ & PA ($\degr$) & $M_V$ & $R_{\mathrm{h}}$ (arcmin) & $R_{\rm proj}$ (kpc) & References\\
    \hline
    And I & $776.2\pm18$ & $-1.51\pm0.02$ & $0.28\pm0.03$ & $30\pm4$ & $-11.4\pm0.2$ & $3.90\pm0.10$ & 44 & 1, 2, 9 \\
    And II & $666.8\substack{+16 \\ -15}$\ & $-1.39\pm0.03$ & $0.16\pm0.02$ & $31\pm5$ & $-11.7\pm0.2$ & $5.30\substack{+0.10 \\ -0.10}$ & 140 & 1, 3, 9\\
    And III & $721.1\substack{+17 \\ -16}$\ & $-1.75\pm0.03$ & $0.59\pm0.04$ & $140\pm3$ & $-9.5\pm0.2$ & $2.00\pm0.20$ & 67 & 1, 2, 9\\
    And VI & $831.8\pm23$ & $-1.50\pm0.10$ & $0.41\pm0.03$ & $163\pm3$ & $-11.3\pm0.2$ & $2.15\pm0.08$ & 268 & 1, 4, 5, 7\\
    And VII & $824.1\pm23$ & $-1.37\pm0.01$ & $0.13\pm0.04$ & $94\pm8$ & $-12.8\pm0.3$ & $3.40\pm0.12$ & 219 & 1, 2, 7\\
    And X & $631.0\substack{+18 \\ -17}$\ & $-2.27\pm0.03$ & $0.10\substack{+0.34 \\ -0.10}$ & $30\substack{+20 \\ -12}$ & $-7.3\pm0.3$ & $1.10\substack{+0.40 \\ -0.20}$ & 76 & 1, 2, 9\\
    And XIV & $772.7\substack{+22 \\ -21}$\ &  $-2.23\pm0.01$ & $0.17\substack{+0.16 \\ -0.17}$ & $-4\pm14$ & $-8.6\pm0.3$ & $1.50\pm0.2$ & 159 & 1, 8, 9\\
    And XV & $748.2\pm17$ &  $-1.43\pm0.42$ & $0.24\pm0.10$ & $38\pm15$ & $-8.4\pm0.3$ & $1.3\pm0.11$ & 92 & 1, 8, 9\\
    And XVI & $517.6\pm19$  & $-2.10\pm0.20$ & $0.29\pm0.08$ & $98\pm9$ & $-7.5\pm0.3$ & $1.59\pm0.16$ & 129 & 1, 6, 9\\
    And XVII & $758.6\substack{+25 \\ -24}$\ & $-1.70\pm0.20$ & $0.50\pm0.10$ & $110\pm9$ & $-7.8\pm0.3$ & $1.40\pm0.30$ & 44 & 1, 4, 9\\
    And XIX & $812.8\substack{+34 \\ -29}$\ & $-1.50\pm0.02$ & $0.58\substack{+0.05 \\ -0.10}$ & $34\pm5$ & $-10.1\pm0.3$ & $14.20\substack{+3.45 \\ -1.97}$ & 104 & 1, 10, 9\\
    And XX & $741.3\substack{+28 \\ -27}$\ & $-2.20\pm0.40$ & $0.11\substack{+0.41 \\ -0.11}$ & $90\substack{+20 \\ -44}$ & $-6.4\pm0.4$ & $0.40\substack{+0.20 \\ -0.10}$ & 125 & 1, 4, 9\\
    \hline
    \end{tabular}
    \caption{Fundamental properties of the M31 dSph sample adopted in this work.
    (1) \citet{Savino2022}; (2) \citet{Kirby2020}; (3) \citet{Ho2012}; (4) \citet{Collins2013}; (5) \citet{Armandroff1999}; (6) \citet{Letarte2009}; (7) \citet{McConnachie2006b}; (8) \citet{Wojno2020}; (9) \citet{Martin2016}; (10) \citet{Cullinane2024}}
    \label{Table:Gal_properties}
\end{table*}

\section{Observations and data reduction}
\label{sample}

\begin{figure*}
	\centering
        \includegraphics[width=2\columnwidth,keepaspectratio]{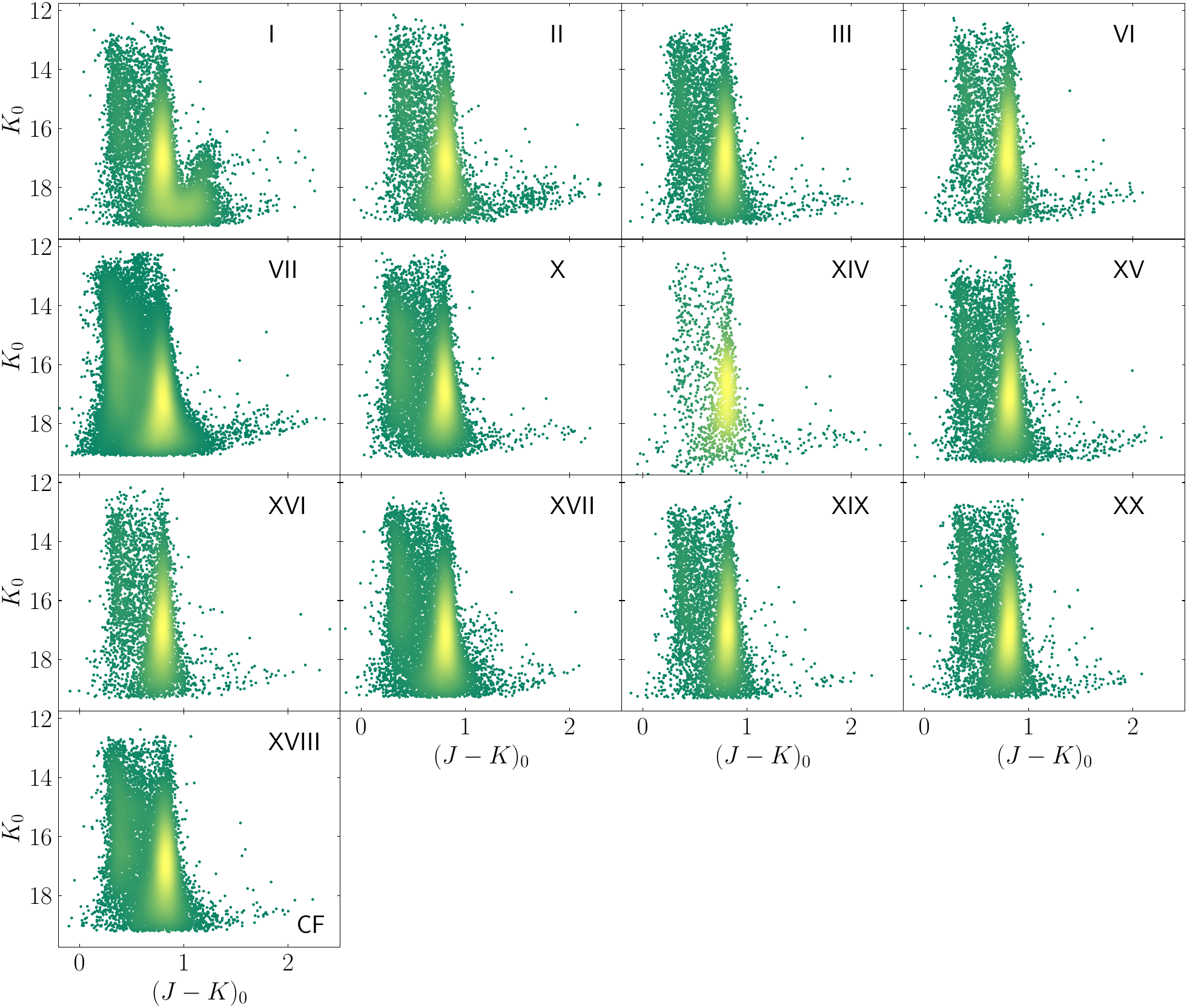}      
	\caption{Extinction-corrected NIR CMDs for the 0.81 deg$^2$ fields listed in Section~\ref{sample}, consisting of 12 dSph fields and the And~XVIII contamination field, displaying all stellar and probably-stellar sources in the WFCAM $J-K$ vs. $K$ bands. The And\,XVIII field is labelled with \enquote{CF} for contamination field. }
\label{CMD_comp}
\end{figure*}

\label{cuts}
\begin{figure}
	\centering
    \includegraphics[width=1\columnwidth,height=0.8\textheight, keepaspectratio]{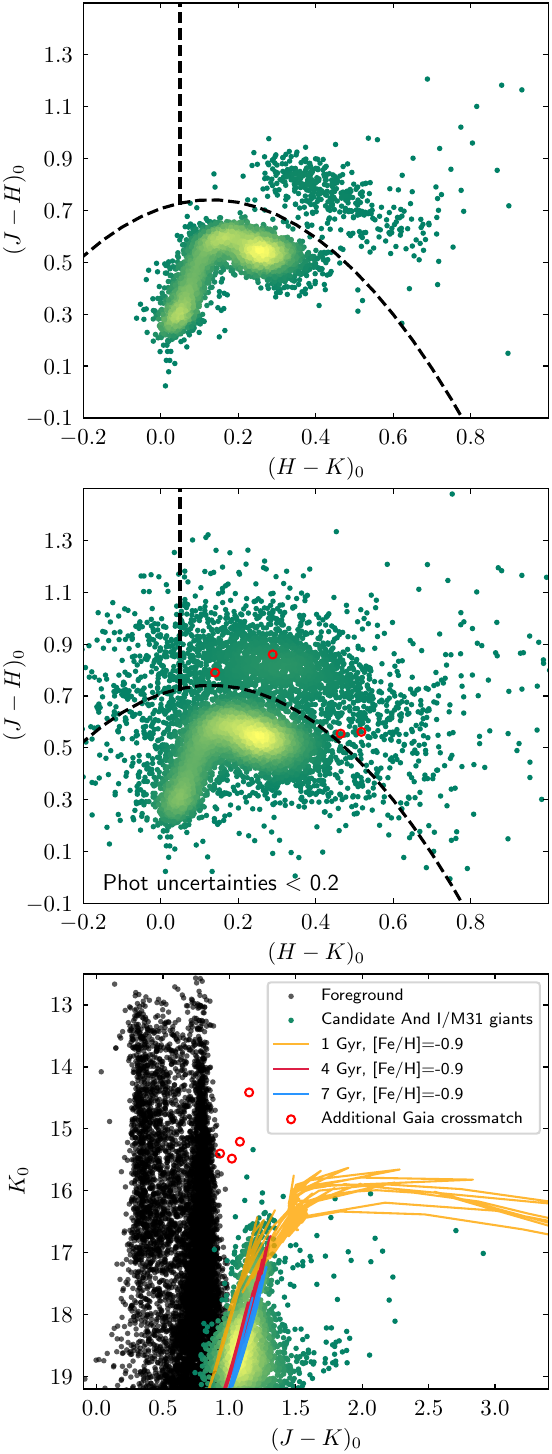}
	\caption{Top: Extinction-corrected $(H-K)_0$ vs. $(J-H)_0$ colour-colour diagram of the And\,I field, indicating the separation of foreground sources from candidate giants (both in the galaxy and the adjacent field), using a clean subsample with photometric errors less than 0.05 magnitudes. The boundaries to remove foreground sources (as well as those with colours too blue to be considered a giant) are overlaid in black. Middle: The same boundaries demonstrated on the full And\,I sample with photometric errors less than 0.2. Bottom: CMD for the same sample as in the middle panel, separated into foreground contaminants and And\,I field giants. Isochrones with a representative metallicity of $\mathrm{[Fe/H]} = -0.9$ and ages of 1, 4 and 7 Gyr are overplotted in orange, red and blue to indicate the expected locations of giants (only the RGB, early-AGB and TP-AGB phases are plotted; the resolution of the thermal pulse cycles has been set to $n_{\rm inTPC} = 20$). We display the foreground dwarf stars with colours redder than our boundaries identified through crossmatching with \emph{Gaia} using red circles in the bottom two panels.}
\label{fore_rem}
\end{figure}

Our dataset was obtained using NIR imaging from with the Wide-Field Camera (WFCAM) on the 3.8\,m UKIRT \citep{Casali2007}, providing coverage not only of the target dSphs but also of the surrounding M31 halo regions. This wide spatial coverage enables a detailed analysis of the AGB populations within the dSphs, as well as those in the adjacent halo fields. The observations were obtained between the 4th and 7th of October 2008 as part of a targeted survey of luminous red stellar populations in Local Group galaxies (PI: Irwin).

The dSphs include: Andromeda I, II, III, VI, VII, X, XIV, XV, XVI, XVII, XIX, and XX. The survey also imaged And\,XVIII, which is not a member of the M31 group and lies at a much larger heliocentric distance ($1178.0\pm43$ kpc), but was included because the distance to this system was not well known at the time of observation. We instead use this field to correct for contamination in the halo regions. The dwarfs span a broad range of properties, with luminosities from $\rm M_V = -12.8$ to $\rm M_V = -6.4$, with the sample dominated by the brighter dSphs. This magnitude range samples most of the luminosity distribution of the M31 dwarf satellites, although it excludes the brightest dIrr and dE systems and the majority of (ultra-faint) dwarfs fainter than about $\rm M_V = -7$. The luminosity distribution of the sample compared to the full M31 dwarf population is shown in Figure~\ref{hist}. 

Square areas around each dSph target were imaged in three NIR wavebands: $J$, $H$, and $K$ \citep{Tokunaga2002}. WFCAM is composed of four $2048\times2048$ Rockwell HgCdTe detectors, with significant separations between the detectors. Four dithered exposures are required to yield contiguous coverage across the square, resulting in a total field of view (FoV) of approximately 0.81 square degrees ($54\arcmin \times 54\arcmin$), with a pixel scale of ${\sim}0.4$ arcseconds per pixel. Each pointing had a total exposure time of 200 seconds in each of the $J$, $H$, and $K$ bands. The average seeing on all frames varied between ${\sim}0\farcs57$--$0\farcs93$ across the dSphs. Photometric uncertainties remain below 0.2 mag for sources brighter than 20.37 -- 21.01 in J, 19.58 -- 20.45 in H, and 19.09 -- 19.82 in K.

The observed dSphs span projected distances from the centre of M31 of approximately 44 to 270\,kpc. Their spatial distribution within the M31 halo is shown in Figure~\ref{pandas_spatial}, alongside the PAndAS survey footprint \citep{McConnachie2018}. The key properties of the dSphs, such as distance, metallicity, and luminosity, are summarised in Table~\ref{Table:Gal_properties}. Distances (column two) are consistently derived from RR Lyrae variable stars \citep{Savino2022}. Column three lists the spectroscopic metallicities derived from Red Giant Branch (RGB) stars \citep{Kirby2020, Ho2012, Collins2010, Wojno2020, Letarte2009}. Structural properties (ellipticity, column four; position angle, column five; half-light radius, column seven) and absolute magnitudes (column six) come from \citet{McConnachie2006b} and \citet{Martin2016}.

The data were processed using the WFCAM pipeline developed by the Cambridge Astronomy Survey Unit (CASU; \citealt{Irwin2004}), which performs standard image reduction steps, including the removal of instrumental signatures and sky background, as well as source extraction. Astrometric and photometric calibration were carried out using the 2MASS point source catalogue. We retained only sources classified as \enquote{stellar} or \enquote{probably stellar} across all bands (corresponding to morphological flags -1 and -2). Sources were further limited to those with photometric uncertainties less than 0.2 mag in all bands. Extinction corrections for each source were then carried out using the Python {\sc dustmaps} package \citet{Green2018}, using the E$(B-V)$ extinction map from \citet{Schlegel1998}, assuming an $R_V$ value of 3.1 and coefficient corrections by \citet{Schlafly2011} of 0.709, 0.449 and 0.302 for $J$, $H$ and $K$, respectively. The average foreground reddening towards the galaxies is $A_J = 0.06$, $A_H = 0.04$, and $A_K = 0.03$.

\section{Selection of AGB stars}   
\label{agbsel}
\subsection{Foreground contamination}
\label{foreground}

In Figure~\ref{CMD_comp}, we present the extinction-corrected $J-K$ versus $K$ Colour-Magnitude Diagrams (CMDs) for the 12 dSph fields and And~XVIII contamination field, following the selection criteria described above. Each field exhibits two or three vertical sequences at $(J-K)_0 \lesssim 1$. These sequences correspond to MW foreground dwarf stars, and their overall and relative densities vary between different fields, reflecting variations in the line-of-sight through the MW. At redder colours, the CMDs also contain the RGB and AGB sequences of the target galaxies. However, in all cases except And\,I, these appear primarily as a diffuse population of faint sources. Background galaxies and red (ultracool) foreground dwarfs also contaminate this region of the CMD.

To address the dominant foreground dwarf star contamination, we used a colour–colour selection designed to isolate the giant stars at the distance of M31. Although the $(J-K)_0$ versus $K_0$ CMD provides some distinction between foreground dwarfs and stars in the target galaxies, the two populations overlap substantially at fainter magnitudes. In contrast, the $(J-H)_0$ versus $(H-K)_0$ diagram offers a more distinct separation between foreground dwarfs and the target galaxy stars, as first demonstrated by \citet{Bessell1988}. As shown in Figure~\ref{fore_rem} (top), the dwarfs form a characteristic L-shaped sequence in the blue quadrant of the diagram, below $(H-K)_0 \simeq 0.5$ and $(J-H)_0 \simeq 0.7$, while giants cluster in the redder part of the diagram, above $(H-K)_0 \simeq 0.3$ and $(J-H)_0 \simeq 0.5$. The example shown in Figure~\ref{fore_rem} is for the field of And\,I, including both the dSph and adjacent field which projects on the Giant Stellar Stream (GSS; \citealt{Ibata2001}), and was chosen due to its high density of C and M stars. We first selected sources with photometric uncertainties below 0.05 (top panel) and used this high-quality subset to define an empirical boundary between the two populations, consisting of a quadratic curve that traces the bend of the L-shaped foreground dwarf sequence. We also introduced a vertical colour cut at $(H-K)_0 = 0.05$, designed to exclude sources that are significantly bluer than the expected colours of AGB and RGB stars. The expected red colours of AGB stars, and separation from the dwarf sequence was highlighted by \citet{Nikolaev2000} for LMC stars observed with 2MASS in NIR colour-colour diagrams (their Figure~2). The resulting boundaries are shown as black dashed lines in the top and middle panels, where the top panel shows the high-quality subset and the middle panel shows all the sources that met our selection criteria in Section~\ref{sample} (i.e., those with photometric uncertainties less than 0.2 mag). The bottom panel shows the separate populations in a CMD diagram; we also overlay three sets of \texttt{PARSEC} stellar isochrones \citep{Bressan2012,chen2015,pastorelli2020} at intermediate ages of 1, 4 and 7 Gyr in orange, red and blue, respectively. The isochrones have a representative metallicity of $\mathrm{[Fe/H]} =-0.9$ , chosen as a compromise between the metallicity of And\,I ($\mathrm{[Fe/H]} =-1.51$) and the adjacent halo field, which is likely to be significantly more metal-rich due to contamination from the GSS \citep{Ibata2001}. The conversion to $\mathrm{[M/H]}$, as the necessary input for \texttt{PARSEC} isochrones, was done using the empirical relation from \citet{Salaris1993} and assuming an $\alpha$-element abundance of $0.38$ as derived for the GSS by \citet{Escala2020}. Only the RGB, early-AGB and TP-AGB phases are plotted, and the resolution of the thermal pulse cycles has been set to $n_{\rm inTPC} = 20$. The isochrones of all three ages generally provide a good match to the stars above the TRGB (visible at $K_0\sim18.1$) while the 4 and 7 Gyr isochrones are consistent with the dominant observed RGB sequence.

Because the fields cover a range of Galactic latitudes, the colours of the foreground stars shift slightly between systems, so we applied small by-eye vertical and horizontal adjustments ($\leq 0.05$ mag in $(J-H)_0$ and $\leq 0.06$ mag in $(H-K)_0$) to the quadratic boundary for each dSph. Although the colour–colour selection removes the majority of foreground contaminants, it targets only the bluer foreground stars. A small number of redder foreground stars may persist, overlapping the RGB and AGB regions. To explore this, we cross-matched our data with the \emph{Gaia} DR3 catalogue \citep{GaiaCollaboration2021} using \texttt{TOPCAT}\footnote{\url{http://www.star.bris.ac.uk/~mbt/topcat/}}, adopting a matching radius of one arcsecond. For sources at the distance of M31, the uncertainties on parallax and proper motion are far larger than the values themselves. We therefore assumed that only local MW foreground sources would have parallax and proper motion measurements larger than 3$\sigma$ from 0 and removed these sources from our galaxy samples. The global parallax bias in Gaia DR3 was corrected according to the recipe laid out in \cite{Lindegren2021b}, and parallax error corrections were applied according to \cite{Fabricius2021}. Applying these parallax and proper motion cuts typically removed between 4--50 additional sources per field. For the And\,I field in particular, four of these additional sources were removed, all of which possess colours akin to an M star, with $(J-K)_0 \sim 0.93$ -- 1.14. These sources are shown as red circles in the bottom two panels of Figure~\ref{fore_rem}. In the case of And\,VII, which was severely affected by foreground contamination, this cross-match to \emph{Gaia} removed a further ${\sim}900$ sources in the field, ${\sim}200$ of which fell within four half-light radii ($R_{\rm h}$). All cross-matched sources lay close to the quadratic boundary defined to remove foreground dwarfs and spanned the full range of K-band magnitudes of the sample.

As an additional check on our removal of foreground contamination, we calculated the number of predicted foreground stars in several of the fields using the {\sc trilegal} model \citep{Girardi2012}. Within the magnitude and colour limits of interest, we found these predictions were generally in good agreement with the number of sources we removed. For example, in And\,I ${\sim}7,600$ were predicted, while we removed ${\sim}7,800$, indicating our method accounts for the expected level of foreground contamination.

\begin{center}
    \begin{figure}
	\includegraphics[width=\columnwidth]{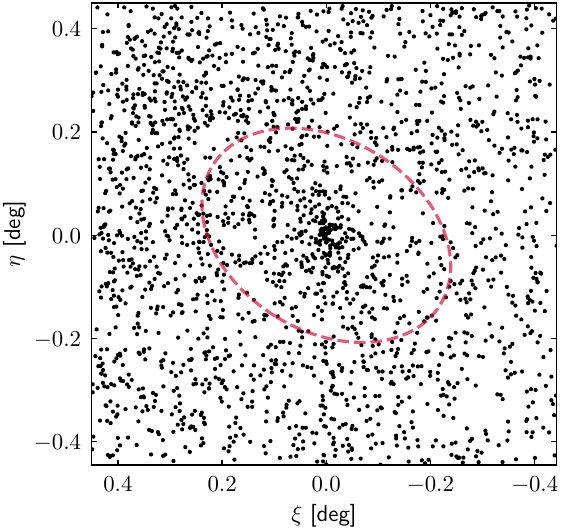}
	\caption{The spatial distribution of stars in the vicinity of Andromeda\,I. A pink ellipse with semi-major axis equal to $4R_{\mathrm{h}}$ (and with the appropriate ellipticity and position angle) is overlaid to illustrate the selection region between Andromeda I stars and M31 halo stars.}
\label{fig:spatial_plot}
\end{figure}
\end{center}
\subsection{TRGB determination} 
\label{TRGB}
The wide FoV of WFCAM allows for the identification of stars associated with both the target dSphs and the surrounding M31 halo. To isolate candidate dSph members, we defined elliptical regions centred on each satellite, with semi-major axes of 4$R_{\mathrm{h}}$, and adopting the corresponding ellipticity and position angle. This is illustrated in Figure~\ref{fig:spatial_plot} for And I, where we overlay the ellipse (in pink) on top of stellar sources in the field, following foreground dwarf decontamination. The large semi-major axis clearly encompasses the full extent of the visible overdensity associated with the dwarf. Sources falling outside these ellipses are assumed to be part of the surrounding halo population, or in this particular case, members of the GSS. The visible gradient in field counts is due to the GSS being denser to the North-East of And\,I.  

Most TP-AGB stars are found above the TRGB, which appears as a drop-off in source density above the RGB. The TRGB can be readily identified using edge detection techniques in galaxies with high stellar densities (see e.g., \citealt{Sibbons2012, Freedman2020}). However, for sparse systems like dSphs, relying on empirical relations between the magnitude of the TRGB and the galaxy metallicity, or the TRGB colour, provides a more robust and reliable approach (see e.g., \citealt{Valenti2004, Gorski2018, bellazzini2004}). In this work, we relied upon the relations between the $(J-K)_0$ colour (in the 2MASS system) and the absolute magnitude of the TRGB, as calibrated in \citet{Freedman2020}, to identify the TRGB in our sample. These relations are based on NIR observations of TRGB stars in the Large Magellanic Cloud (LMC), and have been validated against observations in the Small Magellanic Cloud (SMC) and 11 Galactic globular clusters. In all cases, they are found to agree well. The metallicity range of the LMC, SMC and Galactic globular clusters spans ${\sim}-2.3 < [{\rm Fe/H}] <-0.5$ \citep{VandenBerg2016, Carrera2008}, which captures the metallicity distribution of the dSphs studied here, ensuring that the calibration is appropriate for our sample. The equation given in Section 3.4 of their paper ($M_K = -1.85 \times [(J-K)_0 -1]- 6.14$), along with the adopted distance measurements to the dSphs and colour transformations between 2MASS and WFCAM, are used to estimate the apparent $K$-band magnitude of the TRGB, as follows. 

We first transformed the WFCAM photometry to the 2MASS system using the \citet{Hewett2006} relations. We then determined the average colour of the RGB/AGB sequence in each dSph (within the ellipses defined above) as a proxy for the TRGB $(J-K)_0$ colour in the \citet{Freedman2020} relation, excluding the reddest sources (e.g. C-type AGB stars) by applying an upper $(J-K)_0$ colour cut, which lay approximately between 1.2 and 1.3 depending on the field. The number of stars used to estimate the RGB/AGB colour ranged from ${\sim}10$ to 500. For systems with the smallest samples the derived mean colours are necessarily more uncertain (e.g. And\,X, XIV, XV, and XVII, each with ${\sim}10$ stars). For And\,III, XVI, and XX, the low number of sources ($<10$) within the elliptical regions defined for these dSphs required us to use all sources across the full FoV, including both those inside and outside the elliptical boundaries, to estimate the TRGB. This approach is likely to introduce additional uncertainty due to contamination from non-member stars. We also note that as our colour estimates are derived from stars spanning a broader magnitude range than the exact TRGB, the resulting mean colour may differ slightly from the true TRGB colour, introducing a systematic offset in the inferred TRGB magnitude. To quantify the uncertainty on the TRGB, we performed 1000 Monte Carlo resamplings of the RGB/AGB $(J-K)_0$ distribution, drawing each star from a normal distribution defined by its individual photometric error. For each resampling, we computed the mean colour, and the final TRGB colour was taken as the average of these means, with their standard deviation adopted as the associated uncertainty in the \citet{Freedman2020} relation. The resulting TRGB magnitudes (ranging between 17.53 and 18.44) are listed in Table~\ref{TRGB_locs}. 

To assess the suitability of the \citet{Freedman2020} calibration for our sample, we applied an independent edge-detection method to identify the TRGB in And\,I, the dSph with the most populated RGB. This approach uses the derivative of a smoothed probability density function of the stellar distribution, and is expected to recover the true TRGB in fields with sufficient star counts. The resulting TRGB magnitude was $18.13\pm0.04$, in excellent agreement with the value derived using the \citet{Freedman2020} relation, $18.13\pm0.05$. As such, we consider all sources brighter than the TRGB calculated via the \citet{Freedman2020} relation to be AGB candidates.

Although the dSph TRGB magnitudes vary due to differences in distances and stellar populations, we expect the M31 halo populations around each dwarf to be far more uniform. To facilitate a fair comparison between M31 halo fields, we adopted a fixed M31 halo TRGB of 18.20. This value corresponds to the average TRGB magnitude obtained by the \citet{Freedman2020} relation from the RGB colours of the 12 halo fields, adopting a distance modulus of 24.45 \citep{Savino2022}. This value is also consistent with the location of the TRGB predicted by the \texttt{PARSEC} stellar isochrones. For metallicities in the range $\mathrm{[Fe/H]} = −0.7$ to $−1.5$, representative of the halo between ${\sim}30$ kpc and ${\sim}150$ kpc as reported by \citep{Ibata2014}, and ages between 6 and 9 Gyr, the TRGB magnitude varies from ${\sim}17.9$ to 18.5. 

\begin{figure*}
        \includegraphics[width=2\columnwidth,keepaspectratio]{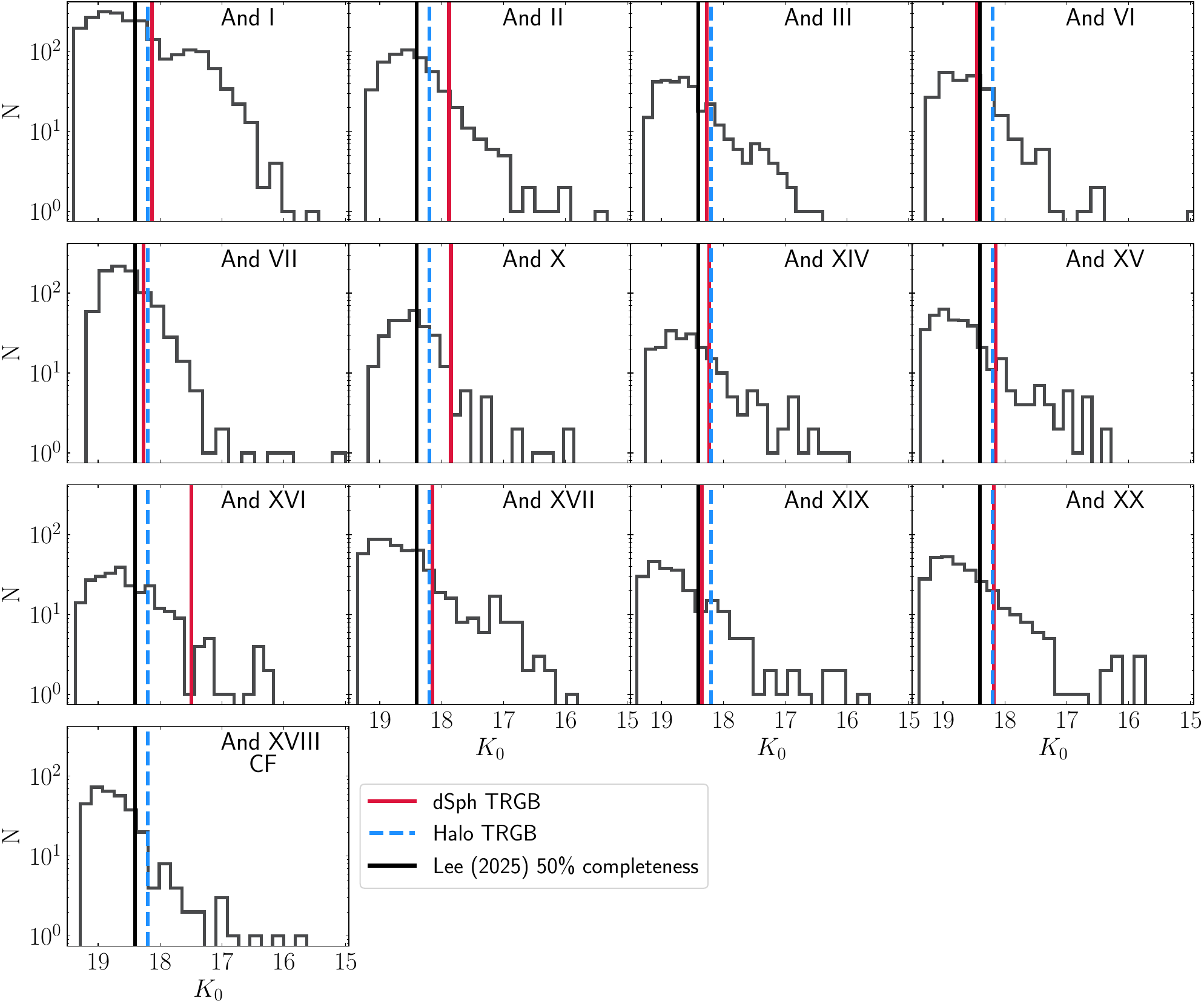}
	\caption{$K_0$-band luminosity functions within each 0.81 square degrees WFCAM field, following foreground contamination removal. The dSph TRGBs (solid red lines) and halo TRGBs (dashed blue lines) are overplotted to demonstrate their proximity to the 50\% completeness limits (solid vertical black lines) taken from \citet{Lee2025}. The And\,XVIII field is labelled with \enquote{CF}, and does not have a dSph TRGB overplotted, as it is designated as our contamination field.}
\label{fig:lum_funcs}
\end{figure*}

\subsection{Completeness}
\label{completeness}

Having established the TRGB magnitudes for both the dSphs and the halo fields, we assess whether the WFCAM data are sufficiently complete at these limits to ensure the AGB stars are reliably sampled. Figure~\ref{fig:lum_funcs} shows the $K$-band luminosity functions for each WFCAM FoV (following foreground star removal; black histograms), together with the apparent TRGB magnitudes derived for each dSph (red solid lines) and their corresponding adjacent halo field TRGBs (blue dashed lines; see Section~\ref{TRGB}). We also plot the 50\% completeness limit from \citet{Lee2025} at $K = 18.4$ mag (solid black line), derived from artificial star tests. The data used by \citet{Lee2025} were obtained as part of the same WFCAM/UKIRT observing programme as our targeted survey data in the M31 halo and were reduced using the same pipeline. They used artificial stars to estimate completeness across a range of stellar densities (0--6, 6--10 and 10--14 stars per arcmin$^2$) and found a consistent 50\% $K$-band completeness limit of 18.4 mag in all cases. Our fields span stellar densities of approximately 2--12 stars per arcmin$^2$, so we adopt this limit for our data. The halo TRGB (18.20) lies above this 50\% completeness limit, and we therefore assume the halo fields are largely complete. The 50\% completeness limit also lies roughly at or below the TRGB magnitude in all dSphs, indicating that the AGB populations in these galaxies are generally well sampled. 

However, as noted above, the TRGB estimation uses the mean colour of stars spanning a range below and above the TRGB as a proxy for the TRGB colour required by the \citet{Freedman2020} calibration. As the RGB/AGB sequence colour varies with magnitude, the resulting mean colour may differ slightly from the true TRGB colour, introducing a systematic uncertainty in the inferred TRGB magnitude. The magnitude range sampled below the TRGB varies between dSphs depending on distance; in the more distant galaxies, where the TRGB lies closer to the 50\% completeness limit, the \citealt{Freedman2020} calibration may bias the inferred TRGB magnitude toward brighter values, as the shallower photometry samples only the brighter portion of the RGB, resulting in a systematically redder mean RGB/AGB colour and a brighter TRGB estimate from the \citealt{Freedman2020}.  We also note that TP-AGB magnitudes vary on timescales of ${\sim}10^2$--$10^3$ days, with some amplitudes exceeding $\Delta K = 1$ magnitude for Mira variables \citep{Jiminez2006, Whitelock2008}. As our observations are single-epoch, the measured magnitudes of the TP-AGBs capture only a snapshot of the brightness variation. This introduces scatter in the observed TP-AGB luminosity function, which can move individual sources across the TRGB boundary and therefore affects the completeness of a snapshot view of the TP-AGB sample.

\subsection{AGB identification}
\label{cm}

\begin{table}
\centering
	\caption{$K_0$-band TRGB magnitudes determined for each of the dSphs.}
	\label{TRGB_locs}
	\begin{tabular}{l c}
	\hline
        \hline
	dSph & $m_{K}^{\mathrm{TRGB}}$\\
	\hline
	And I & $18.13\pm0.05$\\
	And II & $17.88\pm0.05$\\
	And III & $18.27\pm0.1$\\
	And VI & $18.44\pm0.06$\\
	And VII & $18.27\pm0.05$\\
	And X & $17.85\pm0.11$\\
	And XIV & $18.23\pm0.06$\\
	And XV & $18.15\pm0.06$\\
	And XVI & $17.53\pm0.06$\\
	And XVII & $18.17\pm0.10$\\
	And XIX & $18.35\pm0.06$\\ 
	And XX & $18.18\pm0.05$\\
	\hline
	\end{tabular}
\end{table}

\begin{figure*}
	\centering
	\includegraphics[width=2\columnwidth,keepaspectratio]{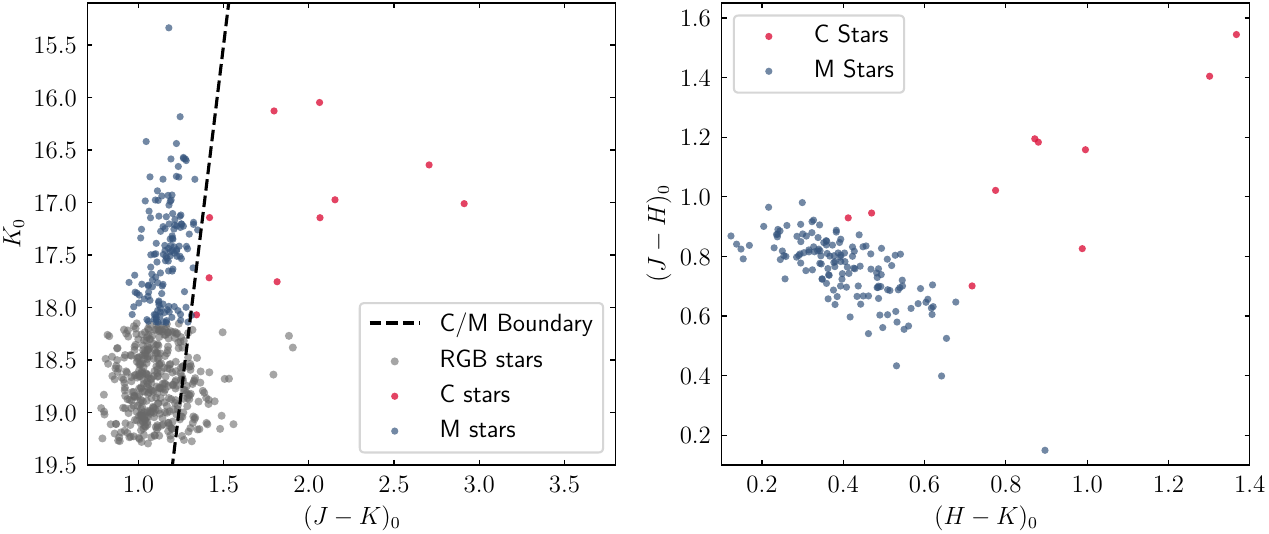}
	\caption{Left: CMD of stellar sources within $4R_{\rm h}$ of And\,I. Red points are C stars, blue points are M stars, and grey points are RGBs. The foreground sources identified with the boundaries described in Section~\ref{agbsel} have been removed. Overplotted in the black dashed line is the C/M star boundary from \citet{Cioni2006}. Right: Colour-colour diagram of AGB sources within And\,I.}
\label{C_M_cuts}
\end{figure*}
 
Having discussed both the TRGB magnitudes and the completeness limits of our dataset, we next isolate and classify the C and M star populations within each dSph. Due to the low mass of these dSphs, identifying features in the CMDs is challenging. As such, the separation between C and M stars was achieved using a cut to the $K_0$ versus $(J-K)_0$ CMD defined by (\citealt{Cioni2006a}, equation 4), and subsequently adopted in various near-infrared studies of nearby galaxies (see e.g., \citealt{Boyer2011} and \citealt{Jones2018}). These cuts were determined using photometry from the DENIS catalogue towards the Magellanic Clouds (DCMC; \citealt{Cioni2000a}) and 2MASS observations of the LMC. In these high stellar density datasets, the AGB sequence and the separation between M- and C-type stars were clearly defined, allowing the boundary to be traced empirically. We note that TP-AGB colours also vary on timescales of ${\sim}10^2$--$10^3$ days, with typical changes of 0.01-0.4 magnitudes, but reaching amplitudes as high as $\Delta(J-K) \sim 1$ for very red, dust-enshrouded C stars. However, a spectroscopic study of the LMC by \citet{Jones2017} confirmed these cuts to be broadly reliable (see their Figure 17, which overlays spectroscopically confirmed C and M stars on the $J-K$ versus $K$ CMD and shows good agreement with the \citet{Cioni2006a} cuts adopted in \citealt{Boyer2011}). Following the approach of \citet{Cioni2006b}, also used in \citet{Boyer2011}, we adjusted the \citet{Cioni2006a} boundary to account for differences in both metallicity and distance between the LMC and our target dSphs. Since metallicity significantly affects the $J-K$ colour of red giant branch stars, we applied the metallicity correction proposed by \citet{Cioni2006b}, using $\Delta (J-K) = 0.118 \times \Delta[\mathrm{Fe/H}]$, which shifted the boundary by ${\sim}0.1$--0.2 mag in $(J-K)_0$. The resulting separation (black dashed line), as well as the locations of candidate C and M stars in both a CMD (left panel) and colour-colour diagram (right panel) are shown in Figure~\ref{C_M_cuts} for And\,I. Visually, the relation provides a good separation between M- and C-type stars, and this separation remains effective across the full dSph sample. 

We crossmatched our AGB candidates within 1 arcsecond to the Hubble Source Catalog (HSC; \citealt{Whitmore2016}) across all dSphs, which combines tens of thousands of source lists from the Hubble Legacy Archive into a single master catalogue. Additionally, we crossmatched to the HST catalogue provided by \citet{Savino2025, Savino2022}, which has photometry derived using Dolphot for all our dSphs.\footnote{The HSC catalogue is available on MAST at \url{https://catalogs.mast.stsci.edu/hsc/}, and the \citet{Savino2025, Savino2022} catalogues are available at \url{https://archive.stsci.edu/hlsp/m31-satellites}.} Because HST has significantly better spatial resolution than our UKIRT/WFCAM images, some sources that appear stellar in our data may be flagged as extended in the HST catalogues. Following \citet{Whitmore2016}, we adopt a Concentration Index threshold of 1.4 to identify and exclude extended sources in the HSC. Two C stars met this criterion across our dSph sample (one in And\,II and one in And\,VII). For the \citet{Savino2025, Savino2022} catalogue, we adopt their cuts on the Round and Sharp$^2$ parameters of 3 and 0.2, respectively, to exclude extended sources. One M star in And\,XIV met these criteria. However, in all three cases, the Concentration Index, Sharp$^2$, and Round values lie very close to the respective thresholds. We therefore retain these objects, but mark them with an asterisk in our final catalogues to reflect this uncertainty.

The first ten C and M-type AGB candidates identified in  And\,I, as an example, are listed in the appendix, in Tables~\ref{table8} and~\ref{table9}. The full sample and classifications are available for each dSph as a machine-readable table with the paper and from VizieR. The numbers of C and M stars found in each dSph are summarised in Table~\ref{dSph_CMs}, with corresponding uncertainties derived from Poisson number statistics.

\begin{table*}
\centering
	\caption{Number of candidate C and M stars, $\mathrm{C}/\mathrm{M}$ ratios (prior to and post halo field subtraction) and derived metallicities in each of the M31 dSphs. Errors are derived from Poisson number statistics and propagating the uncertainties in equation\,\ref{feh_eqn}. 
    Unlike the other dSphs in our sample, the numbers given for And\,XIX are determined within only $2R_{\mathrm{h}}$ because $4R_{\mathrm{h}}$ extends beyond our FoV.}
	\label{dSph_CMs}
        \setlength{\tabcolsep}{5pt}
	\begin{tabular}{@{}l c c c c c c c c c c @{}}
        \hline
        \hline
        & \multicolumn{4}{c}{Pre-subtraction dSph star counts} &  & \multicolumn{4}{c}{Post-subtraction dSph star counts}\\
	dSph & $N_{\mathrm{C}}$ & $N_{\mathrm{M}}$ & $\mathrm{C}/\mathrm{M}$ & $[\mathrm{Fe}/\mathrm{H}]$ & &  $N_{\mathrm{C}}$ & $N_{\mathrm{M}}$ & $\mathrm{C}/\mathrm{M}$ & $[\mathrm{Fe}/\mathrm{H}]$ \\
	\hline
	I & $10\pm3$ & $137\pm12$ & $0.07\pm0.02$ & $-0.86\pm0.14$ && $1\pm1$ & $16\pm4$ & $0.06\pm0.06$ & $-0.82\pm0.24$ \\
    II & $21\pm5$ & $24\pm5$ & $0.88\pm0.26$ & $-1.36\pm0.09$ && $8\pm3$ & $16\pm4$ & $0.50\pm0.21$ & $-1.25\pm0.11$ \\
    III & $3\pm2$ & $1\pm1$ & $3.0\pm3.46$ & $-1.61\pm0.25$ && $2\pm1$ & $0$ & \dots & \dots \\
    VI & $6\pm2$ & $9\pm3$ & $0.67\pm0.35$ & $-1.31\pm0.12$ && $5\pm2$ & $6\pm2$ & $0.83\pm0.50$ & $-1.35\pm0.14$\\
    VII & $31\pm6$ & $30\pm5$ & $1.03\pm0.26$ & $-1.40\pm0.08$ && $13\pm4$ & $9\pm3$ & $1.44\pm0.62$ & $-1.47\pm0.11$ \\
    X & $0$ & $0$ & \dots & \dots && $0$ & $0$ & \dots & \dots \\
    XIV & $2\pm1$ & $5\pm2$ & $0.40\pm0.33$ & $-1.20\pm0.19$ && $1\pm1$ & $3\pm2$ & $0.33\pm0.41$ & $-1.17\pm0.26$ \\
    XV & $0$ & $0$ & \dots & \dots && $0$ & $0$ & \dots & \dots \\
    XVI & $0$ & $0$ & \dots & \dots && $0$ & $0$ & \dots & \dots \\
    XVII & $1\pm1$ & $5\pm2$ & $0.2\pm0.22$ & $-1.06\pm0.24$ && $1\pm1$ & $3\pm2$ & $0.33\pm0.37$ & $-1.17\pm0.24$ \\
    XIX & $4\pm2$ & $27\pm5$ & $0.15\pm0.08$ & $-1.00\pm0.15$ && $1\pm1$ & $7\pm3$ & $0.14\pm0.18$ & $-1.00\pm0.27$ \\
    XX & $0$ & $0$ & \dots & \dots && $0$ & $0$ & \dots & \dots \\
\hline
	\end{tabular}
\end{table*}  

\subsection{Halo field contamination and dSph C/M ratios}
\label{background}
There is likely to be some contamination from the M31 halo within the elliptical regions defined for each dSph, so we apply a correction to account for this and determine a less contaminated estimate of the number of C and M stars genuinely associated with each dSph. This involved calculating the number density of C and M stars both inside and outside the defined elliptical region, using the entire remaining area of the field in the latter case. The number density outside the elliptical region is then subtracted from that inside, separately for both C and M stars, and multiplied by the elliptical area to calculate (to the nearest integer value) the expected number of C and M stars in each dSph. In And\,III, the subtraction led to a negative number of M stars so we instead report zero stars. The $\mathrm{C}/\mathrm{M}$ ratios in each of the dwarfs are then calculated from these corrected numbers and converted to [Fe/H] using the relation from \citet{Cioni2009}, who finds 

\begin{equation}
\label{feh_eqn}
	\mathrm{[Fe/H]} = -1.39\pm0.06 - 0.47\pm0.10 \times \log\left(\frac{\mathrm{C}}{\mathrm{M}}\right).
\end{equation}

\noindent This relation was originally obtained by \citet{Battinellid2005}, who homogeneously classified AGB stars in a wide sample of Local
Group galaxies using the $R$ and $I$-band filters, and the narrowband CN and TiO filters, and calibrated their values against $\mathrm{[Fe/H]}$ from the literature. \citet{Cioni2009} revised this relation using updated metallicity measurements spanning $\mathrm{[Fe/H]} = -0.83 \pm 0.17$ to $\mathrm{[Fe/H]} = -2.08 \pm 0.20$.

In the case of And\,XIX, the 4$R_{\mathrm{h}}$ ellipse extends beyond the WFCAM FoV, with only 58\% of the 4$R_{\mathrm{h}}$ ellipse lying within the FoV. Due to this limitation, we restricted the selection to within 2$R_{\mathrm{h}}$, the largest elliptical region fully contained within the FoV, acknowledging that this likely leads to an underestimate of the total number of AGB stars compared to the other galaxies. Assuming an exponential surface density profile, an aperture of $2R_{\mathrm{h}}$ encloses $\sim$85\% of the total stellar population, compared to $\sim$99\% within $4R_{\mathrm{h}}$.

Table~\ref{dSph_CMs} lists the resulting C/M ratios and corresponding metallicities, with their respective uncertainties derived from Poisson number statistics and propagating the uncertainties in equation\,\ref{feh_eqn}. We list numbers both before and after applying the halo field contamination correction, since these will likely bracket the true number of TP-AGB stars present. In And\,I, as expected, correcting for halo contamination substantially reduces the number of AGB stars, because it coincides with the GSS and therefore has a higher field star density than the other dSph fields. For the dSphs where a C/M ratio could be calculated, the pre- and post-subtraction $\mathrm{[Fe/H]}$ values remain consistent with each other within the uncertainties. 

\subsection{Halo field C/M ratios}
\label{halo_background}
In addition to studying the dSphs themselves, we also make use of the adjacent halo fields to investigate their AGB populations and to trace the metallicity of the M31 halo. While contamination in the dSphs was mitigated using adjacent halo fields, these halo fields can themselves be affected by unresolved background galaxies and residual foreground sources, such as faint ultracool dwarfs, which must be accounted for in our analysis of them. To correct for this, we used the And\,XVIII field, as it lies at a relatively large projected distance (112 kpc; location shown in Figure~\ref{pandas_spatial}), is isolated from any M31 halo substructure and exhibits the lowest overall density of sources in the TP-AGB region in our sample. Furthermore, due to the large distance of And\,XVIII (1178 kpc) and its compact size ($R_{\rm h} \sim 0.9$ arcminutes) we do not expect to detect many, if any, resolved stars from it given the depth of our data. Using the same procedure to identify TP-AGBs as in the dSphs, we determined the number of sources that would be classified as C and M stars in the whole And\,XVIII field. We then calculated the number density of these sources and adopted them as representative of the contamination level. The number density of these background contaminants was subtracted from the TP-AGB star densities in all halo fields. Number densities of C and M stars and $\mathrm{C}/\mathrm{M}$ ratios were then calculated the same way as within the dSphs and listed in Table~\ref{halo_CMs}. And\,XIX is excluded from this analysis due to its large half-light radius encompassing most of the WFCAM FoV.

The densities of TP-AGB stars vary significantly across the halo fields, ranging from 0--100 C stars per square degree and 7--816 M stars per square degree. The highest AGB density occurs in the And\,I halo field at a projected distance of ${\sim}44$ kpc, coincident with the GSS. Specifically, the field contains the largest density of M stars ($n_{\mathrm{M}} = 814 \pm 29$) in the sample. The And\,XVII halo field, which lies at a similar projected distance, also shows an elevated M-star density ($n_{\mathrm{M}} = 117 \pm 11$). Despite its large projected distance (${\sim}220$ kpc), the And\,VII halo field exhibits relatively high densities of both C and M stars. A high C-star density is also observed in the And II\,halo field, which coincides with a MW stream-like feature identified by \citet{Martin2014b}.

\begin{table}
\centering
	\caption{Number density of candidate C and M stars per square degree (after background contaminant subtraction), $\mathrm{C}/\mathrm{M}$ ratios and corresponding metallicities in each of the surrounding M31 halo fields. C and M densities are rounded to the nearest integer, while the $\mathrm{C}/\mathrm{M}$ ratios are calculated using the unrounded values.}
    \label{halo_CMs}
    \begin{tabular}{l c c c c}
    \hline
    \hline
    Halo field & $n_{\mathrm{C}}$ (deg$^{-2}$) & $n_{\mathrm{M}}$ (deg$^{-2}$) & $\mathrm{C}/\mathrm{M}$ & $[\mathrm{Fe}/\mathrm{H}]$ \\
    \hline
    I & $59\pm8$ & $816\pm29$ & $0.07\pm0.01$ & $-0.86\pm0.13$ \\
    II & $98\pm10$ & $7\pm3$ & $14.67\pm8.26$ & $-1.94\pm0.17$ \\
    III & $11\pm3$ & $39\pm6$ & $0.27\pm0.11$ & $-1.12\pm0.12$ \\
    VI & $11\pm3$ & $14\pm4$ & $0.80\pm0.38$ & $-1.34\pm0.11$ \\
    VII & $100\pm10$ & $80\pm9$ & $1.25\pm0.23$ & $-1.44\pm0.07$ \\
    X & $8\pm3$ & $56\pm7$ & $0.14\pm0.06$ & $-0.99\pm0.14$ \\
    XIV & $12\pm3$ & $12\pm3$ & $1.00\pm0.48$ & $-1.39\pm0.12$ \\
    XV &  $0$ & $39\pm6$ & \dots & \dots \\
    XVI & $1\pm1$ & $57\pm8$ & $0.02\pm0.03$ & $-0.62\pm0.33$ \\
    XVII & $1\pm1$ & $117\pm11$ & $0.01\pm0.01$ & $-0.47\pm0.27$ \\
    XX & $0$ & $46\pm7$ & \dots & \dots \\
    \hline
    \end{tabular}
\end{table} 

\section{Discussion}   
\label{discussion}
\subsection{Comparison with previous work}

We identify 316 candidate C and M stars across all dSphs. However, this sample is likely contaminated by M31 halo stars, background galaxies and ultracool dwarfs. After accounting for these contaminants via a local background subtraction, the estimated number of C and M stars is reduced to 92. The numbers for each dSph are listed in Table~\ref{dSph_CMs}.

To place these detections in context, we compare our results with previously identified C and M stars in the dSphs. Past work has detected such stars in nine of the dSphs: And\,I, II, III, VI, VII, X, XIV, XVII, and XIX. These studies have used various identification methods, including spectroscopy, broad-band photometry and narrow-band photometry \citep{Boyer2015,Hamren2016,Kerschbaum2004,Harbeck2004, Cote1999}. It should be noted that the goals of the studies often differed; for example, some surveys targeted only C stars, while others focused on TP-AGB stars without distinguishing between C and M types.

The Dust in Nearby Galaxies (DUSTiNGs) survey \citep{Boyer2015} combined uniform \emph{Spitzer} 3.6 and 4.5 $\itmu$m imaging of nearby galaxies to identify TP-AGBs above the TRGB. These filters are particularly sensitive to warm dust, making them effective at detecting very red TP-AGB stars. The survey fields were narrow, around ${\sim}0.05$ deg$^2$ in each galaxy. While the number of sources above the TRGB was estimated in the overview paper \citep{Boyer2015b}, a catalogue of sources in the M31 dSphs was only provided in the \citet{Boyer2015} paper, which aimed to provide a census of variable, dusty AGB stars, constituting a subset of what would be the total TP-AGB population in these dSphs. They found evidence for TP-AGBs in all of this study's dSphs in \citet{Boyer2015b}, and provided coordinates for variable TP-AGBs in And~I, II, X, XIV, XVII, and XIX in the catalogues from \citet{Boyer2015}. 

\citet{Hamren2016} identified C stars using spectroscopy from the DEIMOS multi-object spectrograph on the Keck\,II 10m telescope. Targets were selected from wide-field imaging or surveys (Mosaic Camera on KPNO, CFHT, and SDSS) across eight dSphs in our sample, although C stars were confirmed only in And\,II and X.

\citet{Kerschbaum2004} identified TP-AGB stars in And\,II using broad-band $V$ and $I$ photometry to determine the TRGB and select candidates based on their $V-I$ colours. Narrow-band filters centred on the TiO and CN molecular bands were also used to classify these TP-AGB stars into C and M stars. Their data covered only a small FoV of $6\farcs5 \times 6\farcs5$. Earlier work by \citet{Aaronson1985b} and \citet{Armandroff1994} also detected C stars in And\,II, although their coordinates were not published. \citet{Cote1999} later spectroscopically confirmed an additional C star below the RGB tip.

Finally, \citet{Harbeck2004} identified C stars in And\,III, VI and VII using the narrow-band CN–TiO method combined with broad-band photometry over a $9\farcs6 \times 9\farcs6$ field, a substantial fraction of which were below the TRGB.

Of the 82 previously reported literature C and M star detections in these studies, 39 were crossmatched within one arcsecond using the full UKIRT/WFCAM dataset before any cuts were applied, while the remaining majority were not recovered. 

To investigate the origin of these unmatched sources, we consider the two primary contributing catalogues, \citet{Kerschbaum2004} and \citet{Boyer2015}. One possibility is that some of these objects are fainter than our detection limit. To test this, we compared the $3.6\itmu$m luminosity functions of the recovered sources from \citet{Boyer2015} with those that were not recovered. The \citet{Boyer2015} stars have [3.6] magnitudes ranging from 13.5 to 19.9. We do not recover any sources fainter than [3.6] $\sim 18$. This suggests that these objects likely fall below the UKIRT/WFCAM detection limit, while remaining detectable in the \emph{Spitzer} mid-infrared imaging. However, this is not seen in the \citet{Kerschbaum2004} sample, where unmatched sources span a similar magnitude range to those recovered. The reason for the unrecovered sources in the \citet{Kerschbaum2004} sample is unclear. They note that the positions in their catalogue may be systematically offset by 0\farcs5–1\farcs2 due to the limited accuracy of their reference stars. Increasing the crossmatch radius to 1\farcs5 recovers 22 additional sources, though some may be chance alignments. Most of these are classified as non-stellar by the WFCAM pipeline or have blue colours consistent with foreground stars.
 
Of the crossmatched sources, 13 of the 39 satisfy our TP-AGB selection criteria and are classified as C or M stars in our sample. A comparison of these C and M stars with the literature detections is presented in Table~\ref{table7}. The table lists the dSph, literature ID, ID from this study, method(s) used to identify candidates, and their classifications in both the literature and this study. We note that one matched source in And\,II is classified as a CH star (also referred to as a dC star) in \citet{Harbeck2004}. This designation denotes carbon stars located below the TRGB, which are thought to originate from mass transfer in compact binary systems rather than from third dredge-up during AGB evolution, and thus do not trace intermediate-age populations \citep{McClure1997}. The apparent discrepancy in classification may stem from differences in the adopted TRGB magnitude; \citet{Harbeck2004} do not specify how the TRGB was determined in their Andromeda dSph analysis. Furthermore, the star is likely to be enshrouded in dust, leading to significantly greater extinction in the optical than in the NIR. As a result, it is plausible that the star appears fainter than the TRGB in the $I$-band CMD of \citet{Harbeck2004}, yet remains above the TRGB in our study due to its brightness in the NIR.

\begin{table*}
\centering
	\caption{AGB star candidates in the Andromeda dSphs from the literature crossmatched to our WFCAM-detected candidates. The literature classification CH refers to C stars that lie below the TRGB.}
	\label{table7}
	\centering
	\begin{tabular}{l c c c c c l}
        \hline
	\hline
	dSph & Lit ID & ID (this study) & Detection & Lit classification & Classification (this study) & Reference \\
	\hline 
        And I & 65699 & andi-c-07 & Mid-IR photometry & x-AGB & C & \cite{Boyer2015} \\
        & 8412 & andi-m-056 & Mid-IR photometry & x-AGB & M & \cite{Boyer2015} \\
        & 18812 & andi-m-067 & Mid-IR photometry & AGB & M & \cite{Boyer2015} \\
        And II & 62376 & andii-c-08 & Mid-IR photometry & x-AGB & C & \cite{Boyer2015} \\
        & 81699 & andii-c-09 & Mid-IR photometry & x-AGB & C & \cite{Boyer2015} \\
        & 131 & andii-m-16 & Spectroscopy & C & M & \cite{Hamren2016} \\
        & C607 & andii-m-05 & Narrow-band photometry & M & M & \cite{Kerschbaum2004} \\
        & C949 & andii-m-09 & Narrow-band photometry & M & M & \cite{Kerschbaum2004} \\
        And VI & and6-6618 & andvi-c-5 & Narrow- and broad-band & C & C & \cite{Harbeck2004}\\
        And VII 
        & and7-5653 & andvii-c-21 & Narrow- and broad-band & C & C & \cite{Harbeck2004}\\
        & and7-4447 & andvii-c-20 & Narrow- and broad-band & CH & C & \cite{Harbeck2004}\\
        & and7-3727 & andvii-m-13 & Narrow- and broad-band & C & M & \cite{Harbeck2004}\\
        & and7-6111 & andvii-m-08 & Narrow- and broad-band & C & M & \cite{Harbeck2004}\\
        \hline
	\end{tabular}
\end{table*}

The comparison in Table~\ref{table7} shows that we recover previously reported oxygen-rich stars as M stars in our study. Likewise, the majority of previously identified extreme AGBs (x-AGBs) --- highly reddened, typically C stars --- are classified as carbon-rich in our sample (three out of four), supporting the robustness of our classification method for these sources.      

However, the comparison also reveals discrepancies between the classification of C stars in the literature and in our study. Three of the five sources identified as carbon-rich in the literature are classified as oxygen-rich in our sample. One of these, spectroscopically confirmed as a C star by \citet{Hamren2016}, lies very close to our adopted NIR colour boundary, underscoring the limitations of using a hard photometric cut for classification. The other two are from \citet{Harbeck2004}, where the stars exhibit relatively blue NIR colours ($(J-K)_0 \leq 1.2$), placing them 0.2–0.3 mag blueward our C/M separation boundary. \citet{Harbeck2004} used narrow-band CN–TiO and broad-band $V-I$ photometry to identify C stars, which more directly trace stellar chemistry. As demonstrated by \citet{Battinelli2007}, who compared the CN–TiO technique with the NIR method, a hard cut in $(J-K)$ versus $K$ is usually appropriate as a red limit for M stars but not always as the blue limit for C stars. Consequently, our approach can lead to the misidentification of some C stars as M stars in these filters, and a small level of C star contamination is therefore likely present in our M star samples. \citet{Battinelli2007} observed this behaviour across three Local Group dwarf galaxies: WLM, IC~10 and NGC~6822, which have corresponding RGB-based metallicities of $\langle\mathrm{[Fe/H]}\rangle=-1.28\pm0.02$ \citep{Leaman2013}, $\mathrm{[Fe/H]}\sim -1.1$ to $-1.2$ \citep{Tikhonov2009, Kim2009}, and $\langle\mathrm{[Fe/H]}\rangle=-0.84\pm0.04$ \citep{Swan2016}. They found that the NIR colours of C stars showed no strong dependence on the metallicity of their parent galaxies, suggesting that any C star contamination introduced by our colour selection is unlikely to vary strongly with galaxy metallicity. Similarly, \texttt{PARSEC} isochrones predict only a small shift in the theoretical blue limit $J-K$ colours of C stars, of order $\sim0.09$ mag, for a 1 Gyr population over the metallicity range covered by our sample ($\mathrm{[Fe/H]}\approx -1.4$ to $-2.2$).

The remaining 26 crossmatched sources did not meet our criteria for TP-AGB classification. We list these in the Appendix Table~\ref{extra_lit_decs}. Of these, 13 were classified as non-stellar or noise-like in the UKIRT/WFCAM pipeline \citep{Irwin2004}, while eight were classified as foreground sources, and four were below the $K$-band TRGB. 

Many of the non-stellar/noise-like crossmatches come from the DUSTiNGs survey (\citealt{Boyer2015}). Four of their sources appeared non-stellar in all passbands, with one additional source classified as non-stellar in the $J$-band only. WFCAM has a higher resolution compared to the InfraRed Array Camera (IRAC) on the \emph{Spitzer} Space Telescope (0\farcs4 vs. 1\farcs22 per pixel). Since these sources are all brighter than the TRGB, it is plausible that WFCAM was able to resolve their extended nature, for example, identifying them as background galaxies, whereas \emph{Spitzer} classified them as point sources. Several other non-stellar sources come from a study of And\,II using the Nordic Optical Telescope (NOT) at La Palma \citep{Kerschbaum2004}. Their typical seeing conditions ranged between 0\farcs8 and 1\farcs22, while WFCAM benefited from slightly improved seeing conditions during the And\,II observations (between 0\farcs86 and 1\farcs05). It may again be possible that the WFCAM pipeline was therefore able to better identify non-stellar sources.  However, we also found non-stellar/noise-like classifications for the CH stars identified in \citet{Harbeck2004}, which had excellent seeing conditions of better than 0\farcs8. Furthermore, a C star from the spectroscopic study by \citet{Hamren2016} matched to a noise-like classification in our study. As such, some C stars may get misclassified in the CASU pipeline. 

The eight sources previously identified as AGB stars but classified as foreground in our study may be due to the strict colour–colour selection we applied (see Section~\ref{foreground}). However, upon checking the colours of these sources, all fell very comfortably within the blue L-shaped foreground dwarf sequence as shown in Figure~\ref{fore_rem}, and not close to our defined boundaries. One also had well-defined parallax and proper motions from \emph{Gaia}. Furthermore, the majority of these foreground dwarfs are crossmatched with \citet{Kerschbaum2004}, who state that their sample of M stars is likely to be highly contaminated by foreground M-type dwarf stars.

In five cases, the literature source is below the TRGB determined from our data. One of these is indeed a dC type --- a dwarf C star that exists below the TRGB. 

Overall, of the nine dSphs with previously reported C/M star detections in the literature, we recover TP-AGB candidates in four systems (And\,I, II, VI and VII), and recover sources in a further four (And\,III, X, XIV and XVII), but which are classified as foreground stars, below the TRGB or non-stellar objects in our catalogue. In the case of And\,XIX, no counterparts are recovered in our catalogue within the adopted cross-matching radius, suggesting that the previously reported candidate is either below our detection limit or spurious. By incorporating the 316 new candidate detections identified in this study, we increase the number of known candidate TP-AGB stars from 82 to 359. This new total excludes the 26 cross-matched literature sources that did not satisfy our TP-AGB classification criteria, but includes previously reported candidates that we did not independently recover, whose classification we therefore cannot update.

\subsection{dSphs}
\subsubsection{C/M ratios}
\begin{figure}
	\centering
	\includegraphics[width=1\columnwidth]{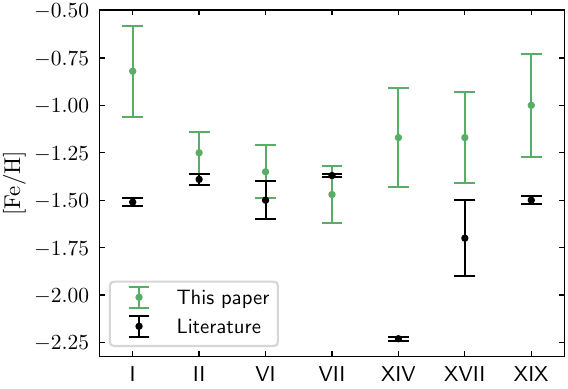}
	\caption{A comparison of adopted spectroscopic metallicities and C/M ratio derived metallicities (post field-contamination subtraction) for the dSph sample.}
\label{fig7}
\end{figure}

\begin{figure*}
	\centering
	\includegraphics[width=2\columnwidth,keepaspectratio]{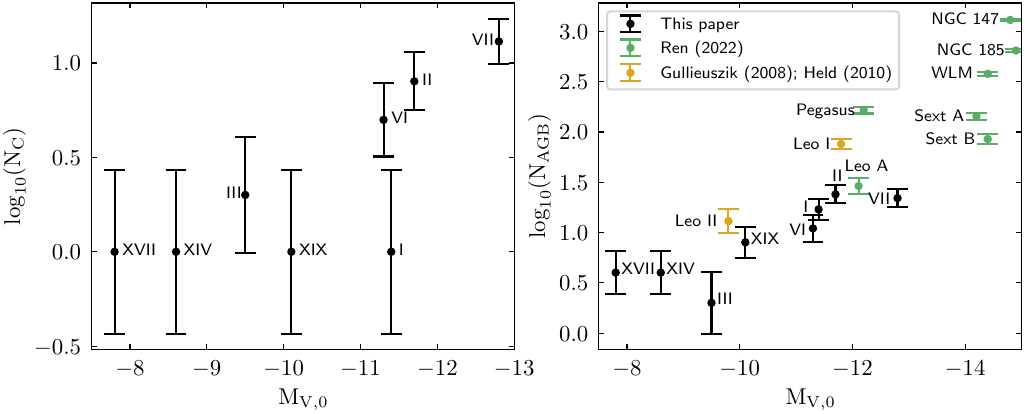}
	\caption{Left: The logarithm of the number of C stars within each dSph vs. the absolute V-band magnitude. The numbers used are post-subtraction of field contamination. Right: The logarithm of the number of all the TP-AGBs vs. absolute V-band magnitude. The data points in gold are from \citet{Gullieuszik2008} and \citet{Held2010} for Leo\,I and Leo\,II, while the green data points are a selection of the dwarfs from \citet{Ren2022}: Leo A, Sextans A, Sextans B, Pegasus dIrr, NGC\,185, NGC\,147 and WLM. }
\label{fig8}
\end{figure*}

Aside from And\,I, the metallicities derived from the C/M ratios (Table~\ref{dSph_CMs}) are generally low, with background-corrected values ranging from $-1.49$ to $-1.0$ dex, falling within the metallicity range used to calibrate the \citet{Cioni2009} relation. These measurements are broadly consistent with literature values within $2\sigma$, although discrepancies remain, particularly for And\,I and And\,XIV. And\,II, VI and VII show closer agreement, within $1\sigma$, as can be seen in Figure~\ref{fig7}. However, most of the dSphs in Figure~\ref{fig7}, with the exception of And\,VII, appear slightly more metal-rich than literature values, which are derived from spectroscopy of RGB stars. 

The interpretation of metallicities inferred from C/M ratios is complicated because the formation of C stars is sensitive to both stellar mass and metallicity \citep[e.g.][]{Karakas2002, Karakas2014}. As mentioned in Section~\ref{intro}, C stars form during the TP-AGB phase when repeated TDU events transport freshly synthesised carbon from the intershell region to the stellar surface. At higher metallicities, the larger initial oxygen abundance means that more carbon must be dredged up before the atmosphere reaches $\mathrm{C/O}>1$, resulting in fewer C stars and lower C/M ratios. In addition, TDU is generally predicted to be less efficient at higher metallicities, further suppressing C star formation. The mass range over which stars become carbon-rich is also restricted and varies with metallicity. At low masses, inefficient TDU and envelope mass loss can prevent sufficient carbon enrichment, while at high masses hot-bottom burning (HBB) can convert dredged-up carbon into nitrogen, inhibiting C star formation. This mass- and metallicity-dependent behaviour is illustrated in Figure 2 of \citet{Straniero2023}, which shows the predicted minimum and maximum initial masses of carbon-star progenitors as a function of metallicity, demonstrating that C stars are more readily produced at lower metallicities and that the progenitor-mass range capable of producing C stars changes with metallicity.

Consequently, the C/M ratio depends not only on metallicity but also on the age distribution of the underlying stellar population, since the limited progenitor-mass range over which stars become carbon-rich corresponds to a limited range of stellar ages. The observed C/M ratio therefore depends on whether a galaxy formed stars at the lookback times associated with efficient C star production \citep[e.g.][]{Sibbons2012,2013ApJ77483B,Sibbons2015,Marigo2017}. These effects are particularly important for the dwarf galaxies in our sample, where the number of observed TP-AGB stars is small (1--13 C stars and 3--16 M stars). Variations in the intermediate-age star formation history or stochastic sampling of the carbon-star progenitor mass distribution can therefore introduce significant scatter in the measured C/M ratios.

The offset observed for And\,XIV is substantially larger than for the other dSphs. One possible explanation is related to the spectroscopic measurements of \citet{Wojno2020}, which were based on relatively shallow data (S/N $\sim$ 7.6 per Angstrom), and which yielded a very low [Fe/H] compared to other dSphs. Their analysis followed the approach of \citet{Yang2013}, in which low S/N spectra from similar stars are co-added to improve the effective S/N. In \citet{Yang2013}, tests using artificially degraded and co-added spectra showed a tendency for the co-added measurements to result in more metal-poor values for intrinsically more metal-rich populations (their Figure 20), although the magnitude of that bias was smaller than the offset we find here. A direct validation of this procedure was not possible for the M31 dwarfs in \citet{Wojno2020}. Instead, they compared the metallicity distributions derived from co-added spectra with those from individual stars (their Figure 8), finding no statistically significant difference. Nevertheless, the individual sample for And\,XIV was small (four stars), and one comparatively metal-rich star ([Fe/H] $\sim -1.5$) lay far outside the metallicity range of the co-added measurements ([Fe/H] $\sim -2.7$ to $-1.9$). Furthermore, And\,XIV has been shown to host a balanced horizontal branch (between red and blue horizontal branch stars \citealt{Martin2017}), which may be more consistent with a higher metallicity than found in \citet{Wojno2020}. Moreover, the discovery paper by \citet{Majewski2007} found that the best-fitting isochrone to the Kitt Peak 4m Mosaic CCD data corresponded to $\mathrm{[Fe/H]} = -1.7$ at a distance modulus of 24.7, while a similarly good fit was obtained with $\mathrm{[Fe/H]} = -1.3$ at a distance modulus of 24.3. The distance modulus adopted in this study, $24.44 \pm 0.06$, from \citet{Savino2022}, lies between these values.

Our results suggest that And\,I has a significantly more metal-rich population than the other dSphs considered here. This is explained by the fact that our measurement ($\mathrm{[Fe/H]} = -0.82\pm0.24$) is more consistent with the GSS on which And\,I projects ($\mathrm{[Fe/H]} \sim -1.0$; \citealt{escala2021elemental-a5a}), indicating that despite our efforts to reduce halo field contamination, our sample is likely still contaminated. To test whether this similarity to the GSS is due to the large size of our adopted galaxy aperture, defined by an ellipse of semi-major axis $4R_{\mathrm{h}}$, we decreased the semi-major axis to $2R_{\mathrm{h}}$ and remade the measurement. We found that the metallicity of And\,I decreased from $\mathrm{[Fe/H]} = -0.82\pm0.24$ to $\mathrm{[Fe/H]} = -1.08\pm0.17$, which brings it closer to the And\,I spectroscopic metallicity of $\mathrm{[Fe/H]} =-1.51\pm0.02$. However, this value remains discrepant at more than $2\sigma$, suggesting that contamination in this region may be significant and difficult to disentangle without spectroscopic information.
We also examined how the smaller ellipse size of $2R_{\mathrm{h}}$ affects the metallicity measurements for all the galaxies shown in Figure~\ref{fig7}. We found that And\,VII shifts significantly by 0.27 dex to $\mathrm{[Fe/H]} = -1.20\pm0.14$, moving it out of $1\sigma$ agreement with the spectroscopic value. The remaining five shift marginally (by $\leq 0.05$ dex), with no preference for moving closer to or further away from the spectroscopic measurement. Both And\,I and And\,VII exhibit substantial internal metallicity spreads (0.34 and 0.36 dex, respectively; \citealt{Kirby2020}), implying that a smaller ellipse preferentially samples the more metal-rich central regions. In conclusion, we assume that our adopted ellipse size is unlikely to bias the metallicity measurements substantially; instead, the larger aperture likely averages over any internal radial metallicity gradients.

We find a clear correlation between the metallicity offset (defined as the difference between metallicities inferred from the C/M ratio and those measured spectroscopically) and galaxy luminosity. In particular, fainter dSphs tend to exhibit larger offsets. This trend likely arises from stochastic fluctuations due to the small number of AGB stars in low-luminosity systems, which strongly affect the observed C/M ratio, and limitations of the C/M–[Fe/H] calibration, which is typically derived from brighter, more massive galaxies and may not accurately represent the behaviour of low-mass, metal-poor systems.

\subsubsection{Star formation histories} 

\begin{figure}
	\centering
	\includegraphics[width=1\columnwidth,keepaspectratio]{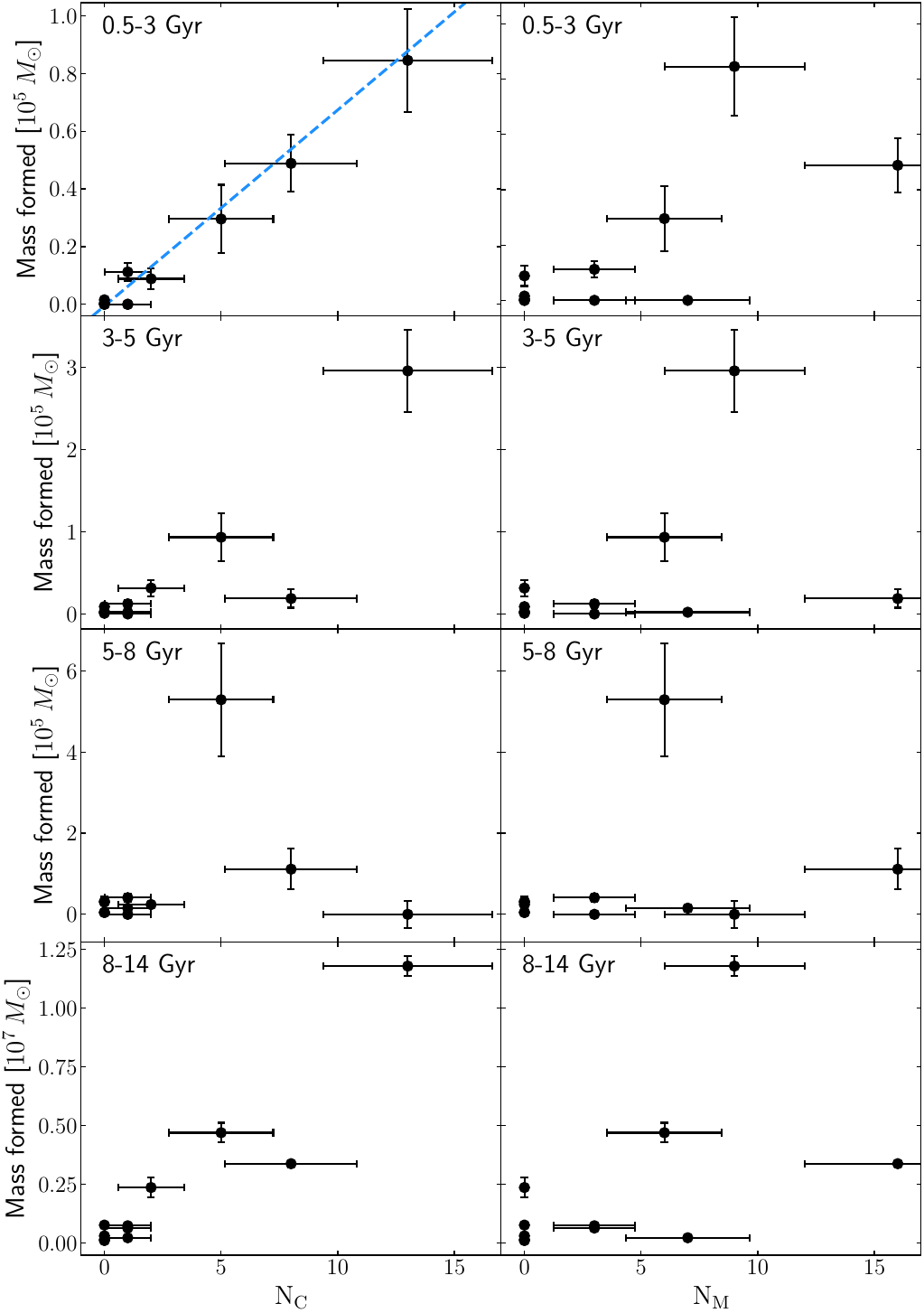}
	\caption{Left: The number of C stars in each dSph versus mass formed during different age bins derived from deep oMSTO CMD fitting by \citet{Savino2025}. Right: The same but for M stars. The blue dashed line in the first panel shows the Bayesian fit to the data. }
\label{C_M_vs_mass}
\end{figure}

Recent studies suggest that the Andromeda dSphs typically experienced prominent early star formation, followed by declining activity and quenching (defined as the lookback time at which 90\% of the total star formation has occurred) approximately 4--9 Gyr ago \citep{Savino2025, Skillman2017, Weisz2019, Abdollahi2023, Saremi2021}. Most M31 dSphs assembled the majority of their stellar mass at relatively early epochs, with little evidence for strong, very recent star formation, unlike MW counterparts such as Fornax and Leo\,I. However, a significant fraction of the dSphs studied in \citet{Savino2025} maintained low-level star formation to intermediate times, quenching around 4--6 Gyr ago. Here, we qualitatively investigate the presence of intermediate-age star formation in our sample from their TP-AGB populations. From the models by \citet{Marigo2017}, the number of C stars in a population is expected to peak in the interval ${\sim}0.5$--3 Gyr (for [M/H] $=−1$ to $−2$) following the burst, while M stars are produced over a broader interval, peaking roughly $0.5$--6 Gyr after a star formation event and persisting to ${\sim}10$ Gyr.

In the left panel of Figure~\ref{fig8}, we demonstrate that our dSphs follow the well-known correlation between the $V$-band luminosity of a galaxy (used as a proxy for stellar mass) and the number of C stars that it harbours (e.g. \citealt{Battinelli2005}). In the right panel, we also show that this relationship holds for the total number of TP-AGB stars (the sum of C and M stars). We also compare our AGB sample to data from similar studies in green and gold \citep{Ren2022,Gullieuszik2008,Held2010}, bearing in mind that the galaxy radii considered are different in each study (for example, the study within Leo\,I used the tidal radius). The comparison studies each use UKIRT/WFCAM infrared data to classify AGB stars but with slightly different methods. \citet{Held2010} and \citet{Gullieuszik2008} (plotted in gold) identify TP-AGB stars in Leo\,I and Leo\,II, selecting AGB stars above the K-band TRGB and separating them into C and M stars in $J-H$ vs $H-K$ colour-colour space. \citet{Ren2022} identify AGB stars in 14 Local Group galaxies, seven of which we include for comparison in green (Leo\,A, Sextans\,A, Sextans\,B, Pegasus dIrr, NGC\,185, NGC\,147 and WLM; these constitute all the dwarfs in their study below $\mathrm{M_V} = -15$). They identified TP-AGB stars above the K-band TRGB and defined cuts between C, M and x-AGB stars in $J-K$ vs $K$. It is clear that these studies follow our trend of increasing AGB counts with absolute V-band magnitude, with no systematic shifts, demonstrating good agreement with our data. 

Leo\,I is known to have experienced strong and recent star formation \citep{RuizLara2021}. At $\mathrm{M_V} = -11.8$, it hosts more AGB stars than other dwarfs of comparable luminosity, lying above And\,II ($\mathrm{M_V} = -11.7$), which quenched ${\sim}7.6$ Gyr ago \citep{Savino2025}. This enhancement could therefore be linked to its recent star formation. However, the picture is not straightforward. Pegasus dIrr ($\mathrm{M_V} = -12.2$; \citealt{Ren2022}) contains substantially more TP-AGB stars than other dwarfs of similar luminosity, such as Leo A ($\mathrm{M_V} = -12.1$) and And\,VII ($\mathrm{M_V} = -12.8$). Yet, both Pegasus and Leo A are reported to have extended star formation histories \citep{Weisz2014}, despite their markedly different AGB counts. Similarly, although we detect no TP-AGB stars in And\,XVI, suggesting little recent star formation, \citet{Skillman2017} find that it has one of the most extended star formation histories among their Andromeda dSph sample. As the numbers of AGBs considered here are all subject to stochasticity in this regime, we refrain from overinterpreting the individual data points but rather note that the overall trend of increasing numbers of AGBs with absolute V-band magnitude holds across the different studies considered. Lastly, although not captured in Figure~\ref{fig8}, we note that the galaxies lacking TP-AGB stars are generally the lowest-mass systems. This trend is consistent with the idea that lower-mass galaxies tend to be preferentially quenched earlier than their higher-mass counterparts (e.g., Figure 1 in \citealt{Weisz2015}). Alternatively, given their extremely low masses, it is also plausible that these systems simply never formed a significant population of TP-AGB stars; in this case, the absence of TP-AGB stars may reflect their intrinsically low luminosities rather than exclusively early quenching.

As C stars trace a narrow age range, we may expect a relationship between the number of C stars a dSph possesses and the mass formed during the epoch of peak C-star production (${\sim}0.5$--3 Gyr). We investigate this using the star formation histories derived from deep old main sequence turn-off (oMSTO) CMD fitting by \citet{Savino2025}, plotting the number of C stars in each dwarf (excluding And\,I due to contamination from the GSS) versus the mass formed in different age bins in Figure~\ref{C_M_vs_mass}. We note that the \citet{Savino2025} SFHs are derived from individual \emph{HST}/ACS fields, which sample only part of each galaxy (between ${\sim}15$--85\% of the total stellar light). Their analysis assumes that the measured SFH is representative of the galaxy as a whole. Any population gradients could therefore introduce additional scatter into the inferred stellar masses.

Despite this potential source of uncertainty, Figure~\ref{C_M_vs_mass} shows a remarkably tight linear relationship between the number of C stars and the stellar mass formed in the 0.5–3 Gyr bin. This correlation weakens for older populations and disappears entirely in the 5--8 Gyr age bin. In contrast, M stars do not exhibit clear trends in any of the age bins, consistent with the theoretical prediction that they reflect stars with a broader age distribution.

Focusing on the 0.5--3 Gyr interval, we performed a Bayesian linear regression of the stellar mass formed 0.5--3 Gyr ago against the number of C stars, accounting for uncertainties in both variables. We obtain the following relation:

\begin{equation} 
\label{Mass_C_eqn} \left(\frac{M_{\rm 0.5-3\,Gyr}}{M_\odot}\right) = (6.8 \pm 2.1)\times10^{3}\,N_{\rm C} - (599 \pm 2120), \end{equation}

where the quoted uncertainties are the posterior $1\sigma$ uncertainties on the fitted parameters.

We note that because the number of C stars is also influenced by metallicity, as well as the dwarfs' SFHs, the relationship likely exists within our sample because the dSphs are all metal-poor (with [Fe/H] between $-1.39$ and $-2.27$) and dominated by early bursts. Over a broader range of metallicities and/or galaxy masses, the relation may become less tight or fail to hold. Nevertheless, it is encouraging that these relations appear when comparing results from narrow \emph{HST} pointings to our wide FoV measurements. 

We observe no obvious trends between the number of TP-AGB stars or the number of C stars vs. the projected radius from the centre of M31. The dSphs without AGB stars tend to cluster at small projected radii, consistent with early quenching via tidal interactions. However, some dSphs at similarly low projected radii still host large AGB populations, suggesting a more complex interplay of environmental and internal processes. Additionally, projected radius is only a two-dimensional proxy for true spatial separation, and does not account for line-of-sight distance.

Overall, the M31 dSphs exhibit diversity in the number of TP-AGBs they possess, with eight of the 12 dSphs possessing AGB stars within $4R_{\mathrm{h}}$, hence showing evidence for intermediate-age star formation. This is consistent with recent studies \citep[e.g.,][]{Weisz2019, Skillman2017, Savino2025}, which demonstrate that many of the M31 dSphs quenched at intermediate-age epochs. Notably, our results also reveal that these eight dSphs also host C stars within $4R_{\mathrm{h}}$, suggesting that star formation likely continued to more recent epochs. Furthermore, the tight relationship between the number of C stars each dSph hosts and the mass formed between ${\sim}0.5$--3 Gyr demonstrates the potential of C stars to estimate the strength of recent star formation in other homogeneously sampled systems, although further testing is needed across a wider sample of galaxies.

\subsection{M31 Halo}  
\begin{figure}
	\centering
	\includegraphics[width=1\columnwidth,keepaspectratio]{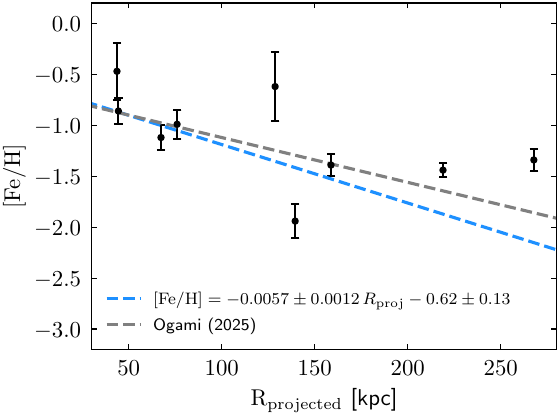}
	\caption{[Fe/H] as derived from the C/M ratios in the M31 halo fields vs. M31-centric projected radius. Also displayed is a least-squares fit to the points within ${\sim} 160$\,kpc (roughly the current survey limit of the halo) in blue and the fit derived from RGB stars in \citet{Ogami2025} in grey. The fits are extrapolated across the full projected distance range of our halo sample (out to the field of And\,VI at 268 kpc).}
\label{fig12}
\end{figure}

We find considerable numbers of TP-AGB stars in most of the stellar halo fields we sample, indicating the presence of non-negligible intermediate-age populations. This is not unexpected given that deep $HST$ star formation history studies have provided evidence for a widespread $\sim2-3$~Gyr population in the halo, extending out to at least 50~kpc \citep{Brown2003db, 2008AJ1351998R,Bernard2015}.  An extended population of TP-AGB stars has also been seen in the inner halo of the nearby spiral NGC\,253 \citep{2014AA562A73G}.

Typical TP-AGB numbers ranged between $24\pm5$ and $875\pm30$ per halo field. There were around ten M stars or more per field, but four of the fields had very few or no C stars (And\,XV, XVI, XVII and XX). Some fields contain high densities of M or C stars. For example, the And\,I halo field, located at a relatively small projected radius (44 kpc), has a notably higher M-star density than the others, as well as one of the highest C-star densities, likely due to contamination from the GSS. Via deep \emph{HST} CMD fitting, \citet{Bernard2015} determined that ${\sim}50\%$ of the GSS stellar mass formed within the last 8 Gyr, supporting the presence of AGB stars. The GSS possesses a high metallicity over the projected radius from M31 where And\,I resides. For example, \citet{Guhathakurta2006} finds $\langle[\mathrm{Fe/H}]\rangle = -0.54$ at a projected radius of 31 kpc, and \citet{escala2021elemental-a5a} report $\langle[\mathrm{Fe/H}]\rangle = -1.03\pm0.07$ over 17--58 kpc from spectra of RGB stars. This high metallicity would be consistent with a high density of M stars (relative to its C stars). There is also a relatively high density of M stars in the field surrounding And\,XVII, which is not associated with any halo streams or features, but lies at a low projected radius (also 44 kpc), where the metallicity in the halo of M31 is relatively high (e.g., see \citealt{Ibata2014, Wojno2023, Gilbert2020}). It is therefore plausible that the high number of M stars in this field is simply due to the fact it samples an inner part of the halo. The cause of the high density of C stars in the And\,II field is unclear. And\,II is located behind the Triangulum–Andromeda (TriAnd) structure, a cloud-like overdensity in the MW (at a heliocentric distance of 16--25 kpc) covering hundreds of square degrees \citep{RochaPinto2004}. Using MW main sequence stars in the PAndAS survey, \citet{Martin2014b} highlighted a wispy, stream-like feature in the TriAnd region where And\,II lies, and estimated it to lie between a heliocentric distance range of 14--20 kpc. We checked to see whether the main sequence stars from this feature could contaminate our sample by placing an isochrone with [M/H] $= -1.5$ and 10 Gyr at the distance of 20 kpc; however, the corresponding stars were too blue to contaminate our sample. And\,VII also shows a modestly elevated AGB density. As it lies at the lowest Galactic latitude in our sample, closest to the plane of the MW, it is more susceptible to residual contamination from Galactic disc stars. Although it seems unlikely that our foreground removal method badly failed to exclude these stars, residual contamination may still contribute to the elevated AGB density.

In Figure~\ref{fig12}, we show the [Fe/H] values derived from the C/M ratios in the halo regions of each WFCAM field (with non-zero counts) as a function of projected distance from the centre of M31. To compare with the currently surveyed extent of M31’s halo (${\sim} 150$ kpc; \citealt{McConnachie2018}), we first performed a linear least-squares fit to the data out to just under 160\,kpc, shown in Figure~\ref{fig12} in blue. A metallicity gradient is apparent in this fit, with a gradient of $-0.0057 \pm 0.0012$ dex kpc$^{-1}$. The fields of And\,XVI and XVII each contain one C star (consistent with zero within the uncertainties) and appear as outliers in the fit. Excluding these points yields a consistent gradient of $-0.0056 \pm 0.0013$ dex kpc$^{-1}$. We note, however, that although the Pearson correlation coefficient suggested a strong association between the variables, the result was not statistically significant (p = 0.1) and likely attributable to the small sample size used in the fit. Nevertheless, our gradient agrees very well with the literature. \citet{Ibata2014} found that the halo metallicity declines from [Fe/H] $=-0.7$ at a projected radius of 27.2 kpc to [Fe/H] $= -1.5$ at 150 kpc (for the full sample including halo substructure, with the smooth halo being ${\sim}0.2$ dex more metal-poor than the full sample at each radius). From our fit, our corresponding values are [Fe/H] = $-0.77 \pm 0.14$ at 27.2 kpc, and [Fe/H] = $-1.47\pm 0.23$ at 150 kpc. From coadded spectra in the SPLASH survey, \citet{Wojno2023} reported a gradient of $-0.0075\pm0.0012$ for the smooth halo over a projected radius of 8--177 kpc, and by fitting the gradient to individual stars, they measured a value of $−0.0053\pm 0.0012$. When including substructure, which necessarily restricted their analysis to a projected radius of ${\sim}90$ kpc where the most distant substructure identified in SPLASH is, they find a slightly steeper gradient of $−0.0086 \pm 0.0023$, but still in agreement. 

When fitting to the full sample, out to ${\sim}270$ kpc (bringing in two additional fields $>200$ kpc), we find a shallower gradient of $-0.0024 \pm 0.0005$ dex kpc$^{-1}$. Although small-number statistics limit firm conclusions, there is a suggestion of a flattening in metallicity beyond ${\sim} 150$ kpc. A qualitatively similar trend is observed in the SPLASH survey, which found that M31’s halo metallicity remains roughly constant beyond ${\sim}100$ kpc \citep{Gilbert2014}. Between roughly 10 and 100 kpc, they measured an inner-halo metallicity gradient of $-0.0110 \pm 0.0007$ dex kpc$^{-1}$, based on photometric metallicities for more than 1500 spectroscopically-confirmed RGB stars \citep{Gilbert2014}. They found a similar metallicity gradient with and without the halo substructure included in their analysis. At face value, our measured slope appears inconsistent with this SPLASH result. However, when restricting their analysis to the radial range 50--90 kpc, \citet{Gilbert2014} derive a significantly shallower gradient of $-0.0048 \pm 0.0047$ dex kpc$^{-1}$. While formally consistent with zero, this best-fit slope is similar to our measurement. Given that our innermost fields lie at roughly 44 kpc, it is plausible that extending our coverage to smaller radii would yield a correspondingly steeper gradient. More recently, \citet{Gilbert2020} derived [Fe/H] for stars in the outer halo (between 43 and 165 kpc in projected radius) via spectral synthesis techniques, finding that the spectral synthesis–based [Fe/H] measurements are largely consistent with the metallicity gradient observed in \citet{Gilbert2014}.

Our measured gradient is also in excellent agreement with the value reported by \citet{Ogami2025}, who find $-0.0044 \pm 0.0004$ dex kpc$^{-1}$ from Dartmouth isochrone fits to RGB star photometry spanning projected radii of ${\sim}$10--120 kpc. They do not exclude halo substructure from their analysis. We also see good agreement between our y-intercept, representing the inferred central metallicity of the stellar halo, and that reported in \citet{Ogami2025}. From our fit out to 160\,kpc, we obtain a y-intercept of $-0.62 \pm 0.13$, which is consistent with the value of $-0.68 \pm 0.03$ reported by \citet{Ogami2025}.

Finally, \citet{Koch2010} performed a spectroscopic analysis of the C/M ratio in fields located in the south-eastern quadrant of the M31 halo, spanning projected radii of 9--160 kpc, and derived an overall value of $0.10 \pm 0.03$. A subset of their pointings specifically targeted the GSS. Their Figure 6 shows that the C/M ratio is essentially flat over 11--32 kpc, the radial range in which they identified C stars (only five were detected in their entire halo sample). Within this interval, a substantial fraction of their fields overlap the GSS. Although their study probes smaller projected radii along the GSS than ours, the And\,I/GSS field at 44 kpc has a C/M ratio of $0.07 \pm 0.01$, in good agreement with their measurement. 

\section{Summary and Conclusions}
\label{conclusion}
We present results from a UKIRT/WFCAM wide-field, NIR photometric study of the TP-AGB populations in 12 of M31's dSph companions as well as its stellar halo. The combined use of colour-magnitude and two-colour diagrams enabled a photometric selection of TP-AGB stars and a separation of these into C and M types. The extent to which the AGB samples are contaminated by populations of foreground MW dwarf stars has been carefully considered and corrected for, while we have accounted for further contamination (e.g., from background galaxies and faint ultracool dwarfs) by using our most sparsely populated and isolated field (And\,XVIII). 

We have derived C/M ratios for the dSphs and adjacent M31 halo fields (surrounding each dSph). Our derived metallicities for the dSphs are broadly consistent with the spectroscopic metallicities of RGB stars published in the literature; while our measurements tend to be somewhat more metal-rich,  they remain within $2\sigma$ in most cases. There is diversity in the number of TP-AGB stars the dSphs host (ranging from zero to 24 following the correction for  contamination) with eight of the 12 dSphs hosting TP-AGBs while the remaining systems host none. The presence of TP-AGB stars in these systems is indicative of intermediate-age star formation, consistent with the results of recent analyses of their star formation histories \citep{Weisz2019, Skillman2017, Savino2025}. These eight systems also host C stars, indicating that star formation continued to more recent epochs (${\sim}0.5$--3 Gyr). Furthermore, a tight linear correlation emerges between the number of C stars in a dSph and the stellar mass it formed between 0.5--3 Gyr as derived from oMSTO fitting, suggesting that C star counts alone can provide a quantitative assessment of the amount of recent star formation in these low-mass systems.

We detect a metallicity gradient in the M31 stellar halo that aligns well with previous studies. Within ${\sim}$160 kpc the gradient is $-0.0057 \pm 0.0012$. The gradient becomes shallower when extended to the halo fields at projected distances of ${\sim}$270 kpc ($-0.0024\pm 0.0005$ dex kpc$^{-1}$). Beyond ${\sim}$150 kpc, there is a suggestion that the gradient flattens, though our sample size in this regime is limited.

Overall, our results reinforce the role of AGB stars as quantitative tracers of intermediate-age stellar populations. Recent work has highlighted how, with sufficient numbers ($\sim1000$) of AGB stars, full modelling of the upper CMD can be carried out to yield reliable star formation histories \citep{Lee2025}. However, in intrinsically low-mass systems such as dSphs, the low numbers of such stars means that this type of modelling will rarely be possible, yet it is precisely these systems that promise critical insight into the processes driving star formation quenching (see the review of \citealt{sales2022baryonic-acc}). 
The empirical correlation we have found between the number of C stars and the stellar mass formed between 0.5--3 Gyr ago in M31 dSphs suggests a possible way to constrain intermediate-age star formation in small galaxies, although further testing will be needed. Exploiting homogeneous, wide-field NIR survey data across the Local Volume, such as that which will be provided by the \emph{Euclid} and \emph{Roman} missions, will be essential for building a more complete and statistically robust census of AGB populations in nearby galaxies and benchmarking their use as robust star formation tracers. Ultimately, this will pave the way for uniform, quantitative analyses of intermediate-age star formation in diverse environments out to $\sim10$~Mpc, a volume in which deep CMDs are either unavailable or unfeasible.

\section*{Acknowledgements}

We thank the referee, Dr Giulia Cinquegrana, for her very helpful suggestions that improved the paper. We also thank Charles Finn for his significant contributions to the initial development of this work.  JH acknowledges support from the Bell Burnell Graduate Scholarship Fund (BB0015), and the School of Physics and Astronomy at the University of Edinburgh. AMNF is supported by UK Research and Innovation (UKRI) under the UK government’s Horizon Europe funding guarantee [grant number EP/Z534353/1] and by the Science and Technology Facilities Council [grant number ST/Y001281/1]. OCJ has received funding from an STFC Webb fellowship. This research made use of the following software, packages and \texttt{python} libraries: \texttt{TOPCAT} (\url{https://www.star.bristol.ac.uk/mbt/topcat}), \texttt{NUMPY} (\citealp{numpy}), \texttt{SCIPY} (\citealp{scipy}), \texttt{ASTROPY} (\citealp{astropy}).

\section*{Data Availability}

The data underlying this article are available in the article and in its online supplementary material. 



\bibliographystyle{mnras}
\bibliography{ref} 




\begin{appendix}
\section{AGB star coordinates}    
\begin{table*}
\centering
	\caption{Coordinates of candidate C stars (prior to background subtraction) alongside $J_0$, $H_0$, $K_0$ magnitudes and absolute $K$-band magnitudes. The example shown is for Andromeda I; the full table will be made available electronically.}
	\label{table8}
	\centering
    \begin{tabular}{l c c c c c c c c}
\hline
\hline
dSph & ID & RA ($\degr$) & Dec ($\degr$) & $J_0$ & $H_0$ &  $K_0$ & $M_{K,0}$ \\
\hline
  And~I & andi-c-01 & 11.30505 & 37.84535 & $18.56 \pm 0.03$ & $17.86 \pm 0.03$ & $17.14 \pm 0.03$ & $-7.31$ \\
  And~I & andi-c-02 & 11.49925 & 37.89596 & $19.57 \pm 0.08$ & $18.74 \pm 0.07$ & $17.76 \pm 0.06$ & $-6.69$ \\
  And~I & andi-c-03 & 11.46482 & 37.96843 & $19.41 \pm 0.07$ & $18.48 \pm 0.06$ & $18.07 \pm 0.07$ & $-6.38$ \\
  And~I & andi-c-04 & 11.61155 & 37.98469 & $18.12 \pm 0.02$ & $16.93 \pm 0.02$ & $16.05 \pm 0.01$ & $-8.40$ \\
  And~I & andi-c-05 & 11.64546 & 37.99973 & $19.13 \pm 0.05$ & $17.97 \pm 0.04$ & $16.98 \pm 0.03$ & $-7.47$\\
  And~I & andi-c-06 & 11.41489 & 38.02381 & $19.35 \pm 0.06$ & $17.94 \pm 0.03$ & $16.64 \pm 0.02$ & $-7.81$\\
  And~I & andi-c-07 & 11.38821 & 38.05255 & $19.92 \pm 0.11$ & $18.38 \pm 0.05$ & $17.01 \pm 0.03$ & $-7.44$ \\
  And~I & andi-c-08 & 11.55145 & 38.10229 & $19.22 \pm 0.05$ & $18.02 \pm 0.03$ & $17.15 \pm 0.03$ & $-7.30$\\
  And~I & andi-c-09 & 11.23062 & 38.1245 & $17.93 \pm 0.02$ & $16.90 \pm 0.02$ & $16.13 \pm 0.01$ & $-8.32$\\
  And~I & andi-c-10 & 11.23245 & 38.12456 & $19.13 \pm 0.06$ & $18.19 \pm 0.05$ & $17.72 \pm 0.06$ & $-6.73$\\
  \hline
	\end{tabular}
\end{table*}

\begin{table*}
\centering
	\caption{Coordinates of candidate M stars (prior to background subtraction) alongside $J_0$, $H_0$, $K_0$ magnitudes and absolute $K$-band magnitudes. The example shown is for the first 10 candidates within Andromeda I; the full table will be made available electronically.}
	\label{table9}
	\centering
\begin{tabular}{l c c c c c c c c}
\hline
\hline
dSph & ID & RA ($\degr$) & Dec ($\degr$) & $J_0$ & $H_0$ &  $K_0$ & $M_{K,0}$ \\
\hline
  And~I & andi-m-001 & 11.39755 & 37.84144 & $18.53 \pm 0.03$ & $17.89 \pm 0.03$ & $17.21 \pm 0.03$ & $-7.24$\\
  And~I & andi-m-002 & 11.45896 & 37.85156 & $18.70 \pm 0.04$ & $18.00 \pm 0.04$ & $17.42 \pm 0.04$ & $-7.03$\\
  And~I & andi-m-003 & 11.34656 & 37.85604 & $18.85 \pm 0.04$ & $18.02 \pm 0.04$ & $17.79 \pm 0.05$ & $-6.66$\\
  And~I & andi-m-004 & 11.36308 & 37.85651 & $18.68 \pm 0.03$ & $17.86 \pm 0.03$ & $17.52 \pm 0.04$ & $-6.93$\\
  And~I & andi-m-005 & 11.35199 & 37.85996 & $18.08 \pm 0.02$ & $17.42 \pm 0.02$ & $16.98 \pm 0.03$ & $-7.47$\\
  And~I & andi-m-006 & 11.37443 & 37.86042 & $18.85 \pm 0.04$ & $18.05 \pm 0.04$ & $17.65 \pm 0.05$ & $-6.80$\\
  And~I & andi-m-007 & 11.24228 & 37.86722 & $19.37 \pm 0.06$ & $18.45 \pm 0.05$ & $18.12 \pm 0.07$ & $-6.33$\\
  And~I & andi-m-008 & 11.33119 & 37.86774 & $18.35 \pm 0.03$ & $17.76 \pm 0.03$ & $17.34 \pm 0.04$ & $-7.11$\\
  And~I & andi-m-009 & 11.48479 & 37.87149 & $18.83 \pm 0.04$ & $17.96 \pm 0.04$ & $17.65 \pm 0.05$ & $-6.80$\\
  And~I & andi-m-010 & 11.16907 & 37.87764 & $18.50 \pm 0.03$ & $17.94 \pm 0.03$ & $17.39 \pm 0.04$ & $-7.06$\\
\hline\end{tabular}
\end{table*}

\begin{table*}
\centering
	\caption{AGB star candidates in the Andromeda dSphs from the literature cross-matched to UKIRT/WFCAM detections that did not make our final TP-AGB sample. The literature classification CH and dC refers to C stars that exist below the TRGB.}
	\label{extra_lit_decs}
	\centering
	\begin{tabular}{l c c c c c c c l}
        \hline
	\hline
	dSph & Lit ID & RA (deg) & DEC (deg) & Detection & Classification & Classification (this study) & Reference \\
	\hline
        And~I & 58559 & 11.40 & 38.11 & Mid-IR photometry
 & x-AGB & Non-stellar in J-band & \cite{Boyer2015} \\ 
        And~II 
        & 30 & 19.1103 & 33.4300 & Spectroscopy & AGB & Foreground & \cite{Cote1999} \\
        & C380 & 19.0885 & 33.4229 & Narrow-band photometry & M & Non-stellar in J and H & \cite{Kerschbaum2004} \\
        & C398 & 19.0892 & 33.3743 & Narrow-band photometry & M & Non-stellar in all bands & \cite{Kerschbaum2004} \\
        & C884 & 19.11 & 33.40 & Narrow-band photometry & M & Non-stellar in all bands & \cite{Kerschbaum2004} \\
        & C1183 & 19.1183 & 33.4073 & Narrow-band photometry & C & Non-stellar in J and K & \cite{Kerschbaum2004} \\
        & C1083 & 19.1149 & 33.4250 & Narrow-band photometry & C & Non-stellar in H and K & \cite{Kerschbaum2004} \\
        & C647 & 19.0993 & 33.3891 & Narrow-band photometry & M & Foreground & \cite{Kerschbaum2004} \\
        & C563 & 19.0959 & 33.3896 & Narrow-band photometry & M & Below TRGB & \cite{Kerschbaum2004} \\
        & C525 & 19.0943 & 33.3970 & Narrow-band photometry & M & Foreground & \cite{Kerschbaum2004} \\
        & C72 & 19.0710 & 33.4186 & Narrow-band photometry & M & Foreground & \cite{Kerschbaum2004} \\
        & C1195 & 19.1186 & 33.3746 & Narrow-band photometry & M & Below TRGB & \cite{Kerschbaum2004} \\
        & C1515 & 19.1316 & 33.3911 & Narrow-band photometry & M & Foreground & \cite{Kerschbaum2004} \\
        & C1237 & 19.1208 & 33.3807 & Narrow-band photometry & M & Below TRGB & \cite{Kerschbaum2004} \\
        & C1055 & 19.1136 & 33.4083 & Narrow-band photometry & M & Below TRGB & \cite{Kerschbaum2004} \\
        & C1442 & 19.1284 & 33.4160 & Narrow-band photometry & M & Foreground & \cite{Kerschbaum2004} \\
        & C387 & 19.0888 & 33.4220 & Narrow-band photometry & M & Foreground & \cite{Kerschbaum2004} \\
        & C2123 & 19.1717 & 33.4450 & Narrow-band photometry & M & Foreground & \cite{Kerschbaum2004} \\
        And~III & and3-2538 & 8.8968 & 36.4836 & Narrow- and broad-band & dC & Below TRGB & \cite{Harbeck2004}\\
        And~VI & and6-10813 & 357.9957 & 24.5797 & Narrow- and broad-band & CH & Non-stellar in K-band & \cite{Harbeck2004}\\
        And~VII & and7-4002 & 351.6321 & 50.6827 & Narrow- and broad-band & CH & Noise-like in K-band & \cite{Harbeck2004}\\
        And~X & 1000019 & 16.6468 & 44.8018 & Spectroscopy & C & Noise-like in K-band & \cite{Hamren2016}\\
        & 70901 & 16.6130 & 44.8194 & Mid-IR photometry & AGB & Non-stellar in all bands & \cite{Boyer2015}\\
        & 98807 & 16.5393 & 44.7976 & Mid-IR photometry & AGB & Non-stellar in all bands & \cite{Boyer2015}\\
        And~XIV & 71257 & 12.8500 & 29.6774 & Mid-IR photometry & AGB & Non-stellar in all bands & \cite{Boyer2015}\\
        And~XVII & 67757 & 9.2517 & 44.3333 & Mid-IR photometry & x-AGB & Non-stellar in all bands & \cite{Boyer2015}\\
	\hline
	\end{tabular}
\end{table*}

\end{appendix}

\bsp	
\label{lastpage}
\end{document}